\documentstyle[pra,aps,preprint]{revtex}
\begin{document}

\title{EXOTIC LOW DENSITY FERMION STATES IN THE TWO MEASURES FIELD THEORY:
  NEUTRINO DARK ENERGY}

\author
{E. I. Guendelman \thanks{guendel@bgumail.bgu.ac.il} and A.  B.
Kaganovich \thanks{alexk@bgumail.bgu.ac.il}}
\address{Physics Department, Ben Gurion University of the Negev, Beer
Sheva 84105, Israel} \maketitle

\begin{abstract}
There exist field theory models where the fermionic
energy-momentum tensor contains a term proportional to
$g_{\mu\nu}\bar{\psi}\psi$ which may contribute to the dark
energy. We show that this new field theory effect can be achieved
in the Two Measures Field Theory (TMT) in the cosmological
context. TMT is an alternative gravity and matter field theory
where the gravitational interaction of fermionic matter is reduced
to that of General Relativity when the energy density of the
fermion matter is much larger than the dark energy density. In
this case also the 5-th force problem is solved automatically. In
the opposite limit, where the magnitudes of fermionic energy
density and scalar field dark energy density become comparable,
nonrelativistic fermions can participate in the cosmological
expansion in a very unusual manner. Some of the features of such
Cosmo-Low-Energy-Physics (CLEP) states are studied in a toy model
of the late time universe filled with homogeneous scalar field and
uniformly distributed nonrelativistic neutrinos, and  the
following results are obtained: neutrino mass increases as
$m\propto a^{3/2}$ ($a$ is the scale factor); the proportionality
factor in the non-canonical contribution to the neutrino
energy-momentum tensor (proportional to the metric tensor)
approaches a constant as $a(t)\rightarrow\infty$ and therefore the
non-canonical contribution to the neutrino energy density
dominates over the canonical one $\sim m/a^3\sim a^{-3/2}$ at the
late enough universe; hence the neutrino gas equation-of-state
approaches $w=-1$, i.e. neutrinos in the CLEP regime behave as a
sort of dark energy
 as $a\rightarrow\infty$; the equation-of-state for the
total (scalar field$+$neutrino) energy density and pressure also
approaches $w=-1$ in the CLEP regime; besides the total energy
density of such universe is less than it would be in the universe
filled with the scalar field alone. An analytic solution is
presented. A domain structure of the dark energy seems to be
possible. We speculate that decays of the CLEP state neutrinos may
be both an origin of cosmic rays and responsible for a late
super-acceleration of the universe.  In this sense the CLEP states
exhibit  simultaneously new physics at very low densities and for
very high particle masses.

\end{abstract}

\section{Introduction}

In the light of the evidence that our universe is
accelerating\cite{accel}, one of the most fundamental questions
facing modern physics is the nature of the dark energy\cite{d.e.}.
In addition, the problem of the nature of the dark matter as well
as the puzzling fact that  at the present time the dark matter and
the dark energy densities appear to be of the same order of
magnitude, appear as a complete mystery. In the absence of a
fundamental theory describing the nature of the dark energy and
the dark matter, this observable "cosmic coincidence"\cite{coinc}
 makes it natural to assume that there should be a
certain dynamical correlation or even a coupling between the dark
energy and the dark matter. Having this in mind, in the context of
quintessence scenarios\cite{quint} of an accelerating expansion
for the present day universe a number of promising approaches have
been developed in order to explain this observable fact, for
example coupled quintesence\cite{amend}, variable mass particles
models\cite{vamp} and different modifications of these and similar
ideas\cite{diff}.

Such modifications of the particle physics theory have a number of
problems (see for example\cite{prob}). One of the most fundamental
problems is the following: although there are justifications for
coupling of the quintessence scalar $\phi$ to dark matter in the
effective Lagrangian, it is not clear why similar coupling to the
baryon matter is absent or essentially suppressed (such coupling
would be the origin of a long range scalar force\cite{longrange}
because of the very small mass of $\phi$). This "fifth force"
problem might be solved\cite{carroll} if there would be a shift
symmetry $\phi\rightarrow\phi + const$. But the quintessence
potential itself does not possess this symmetry although at the
present epoch it must be very flat\cite{kolda}.

Specific properties of neutrinos (which constitute a fraction of
the dark matter) served as a basis for a number of
models\cite{Neutr-dark} concerning a possible coupling of
neutrinos to the dark energy. Recently a new idea has been
suggested\cite{FNW-1} (see also \cite{FNW-2}) that coupling of the
neutrinos to a scalar allows to formulate the effective picture
where the dark energy density depends on the neutrino mass
(treated as a dynamical field). In this model the observable
equation-of-state for the dark energy $w\approx -1$ is obtained if
one assumes\cite{FNW-1} that the neutrino mass $m_{\nu}$ depends
on the density of the background nonrelativistic neutrinos
$n_{\nu}$ in the following way: $m_{\nu}\propto n_{\nu}^{w}$.
However, in the model \cite{FNW-1} the energy density in neutrinos
is small compared to the energy density in the total dark energy
sector.

In this paper we show that in the framework of the Two Measures
Field Theory (TMT) one can address the above goals while
automatically solving some of the mentioned difficulties. For
example, in spite of the presence of the scalar field $\phi$
potentials, the model allows a scale symmetry which involves the
shift transformation $\phi\rightarrow\phi + const$. Some other
results of our TMT model are the following:

1. The shift symmetry $\phi\rightarrow\phi + const$ gets
spontaneous breakdown and as a consequence this results in the
appearance of a dimensionful parameter.

2.  Generically fermions in TMT are very much different from what
one is used to in normal field theory.  For example the fermion
mass can depend upon the fermion  density. Only if the local
energy density of the fermion is much larger than the vacuum
energy density, the fermion can have a constant mass. However this
is exactly the case of atomic, nuclear and particle physics,
including accelerator physics and high density objects of
astrophysics. This is why to such "high density" (in comparison
with the vacuum energy density) phenomena we will refer as "{\it
normal particle physics conditions}" and the appropriate fermion
states in TMT  we will call "{\it regular fermions}". For generic
fermion states in TMT we will use the term "{\it primordial
fermions}" in order to distinguish them from regular fermions.

3.  One of the differences of the gravitational equations from
those of the Einstein's General Relativity (GR) consists in the
appearance of a noncanonical term in the fermion energy-momentum
tensor $\propto g_{\mu\nu}\bar{\psi}\psi$. There exists also an
effective Yukawa coupling of the dilaton scalar field $\phi$ to
{\it primordial} fermions that may be an origin of a long-range
scalar force. However, the coupling of {\it regular} fermions to
gravity approximates very closely that of GR with the correction
being suppressed by factor of $\rho_{d.e.}/\rho_{f}$, where
$\rho_{d.e.}$ and $\rho_{f}$ are the dark energy and local fermion
energy densities respectively. The same is true also for the 5-th
force suppression, i.e for the effective Yukawa coupling of
regular fermions to the scalar field $\phi$.

4. In the opposite regime, where the local energy density of the
primordial fermion is comparable with the vacuum energy density,
TMT predicts the possibility of exotic states whose long range
interactions are very much different from GR because of their
noncanonical coupling to gravity and non-suppressed effective
Yukawa coupling to scalar field $\phi$. In other words, for the
fermion matter in such exotic states GR is not applicable.

As an example of such exotic states, in this paper we study
solutions in a toy model of the late time universe filled with
homogeneous scalar field $\phi$ and uniformly distributed
nonrelativistic neutrinos. The following results are obtained as
the local energy density of the neutrino becomes comparable with
the vacuum energy
 density: neutrino mass increases as $m\sim
a^{3/2}$ ($a$ is the scale factor); the non-canonical contribution
to the neutrino energy-momentum tensor $\propto
g_{\mu\nu}\bar{\psi}\psi$ dominates over the canonical one $\sim
m/a^3$ and as a consequence the neutrino energy density scales as
a sort of dark energy with equation-of-state approaching $w=-1$ as
$a\rightarrow\infty$; the total (scalar field$+$neutrino) energy
density scales as a sort of dark energy and approaches a constant
as $a\rightarrow\infty$; besides the total energy density of such
universe is less than it would be in the universe free of
fermionic matter at all. The latter means that in the framework of
our toy model, regular nonrelativistic neutrino should undergo
transition to an  exotic state described above which we will
denote Cosmo-Low-Energy-Physics (CLEP) state. Demonstration of
this {\it new field theory effect} in the context of TMT is the
purpose of this paper.

However before doing this in a systematic way in TMT, we show in
Sec.2 that there is a possibility of obtaining a fermion
contribution into the dark energy in the context of standard field
theory. However, as we will explain, this attempt is not
successful. In sections 3 and 4 we briefly review general ideas of
TMT and formulate our scale and gauge invariant model. In Sec.5 we
shortly discuss results for dark energy in the absence of
fermions. Sec.6 contains a brief review of regular fermions in TMT
and closely related to them question of reproducing GR. Sec.7
deals with the main subject of the paper: possible reorganization
of the dark energy as a result of the appearance of the CLEP state
neutrinos.

\section{Attempts to Obtain Fermionic Dark Energy in the Context
of Standard Field Theory}

In GR the free fermion part of the action has  the form
 $S^{(f)}_{0}=\int d^{4}x \sqrt{-g}L_{f}$ where $L_{f}$ is
 the free fermion Lagrangian density in the Einstein-Cartan
 space-time. Then a potential cosmological-like term
 $\propto g_{\mu\nu}L_{f}$ naively appears but the Dirac equation
 forces $L_{f}$
 to vanish.

Here we would like to give some feeling of the mechanism by which
 a term in the energy-momentum tensor proportional to
 $g_{\mu\nu}f(\bar{\psi}\psi)$ (where $f(\bar{\psi}\psi)$ is a function of
 $\bar{\psi}\psi$) could be
 obtained in  the framework of the standard (non TMT) field
 theories. Such a term may cause a negative fermion contribution to the
 pressure and therefore might play an important role for the total
 dark energy density. We give in this section two examples of
 models where on the classical level the fermion contributes
 a term in the energy-momentum tensor proportional to
 $g_{\mu\nu}f(\bar{\psi}\psi)$.

 \subsection { A model with coupling to topological density}

 If  the action includes a topological density then the
 term proportional to
 $g_{\mu\nu}\bar{\psi}\psi$ can appear in the energy-momentum
 tensor. As an  example of a topological density we consider here
 $\Omega\equiv\epsilon^{\alpha\beta\mu\nu}Tr(F_{\alpha\beta}F_{\mu\nu})$
 in a gauge theory.

 For illustration let us consider a model where in addition to
 $L_{f}$, which includes now a minimal coupling to gauge fields,
 there is also the interaction of the fermion to the scalar field
 $\phi$ which in its turn couples to the topological density
 $\Omega$ (the latter is parity violating):
 \begin{equation}
 S_{1}=\int \sqrt{-g}d^{4}x\left[\frac{1}{2}g^{\mu\nu}\phi_{\mu}\phi_{\nu}
 -\frac{1}{2}m_{\phi}^{2}\phi^{2}+\lambda_{f}\phi\bar{\psi}\psi\right]
 +\int \lambda_{top}\phi\Omega d^{4}x.
 \label{2-1-L1}
\end{equation}
By integrating out the $\phi$-field one can get (assuming that
$\phi$ is slowly varying) the following fermion part of the
effective action
\begin{equation}
S^{(ferm)}_{eff}=\int\sqrt{-g}d^{4}x\left[L_{f}+
\frac{\lambda_{f}^{2}}{2m_{\phi}^{2}}(\bar{\psi}\psi)^{2}\right]
+\int\frac{\lambda_{f}\lambda_{top}}{m_{\phi}^{2}}\Omega\bar{\psi}\psi
d^{4}x. \label{2-1-L2}
\end{equation}
Now the fermion equation yields
\begin{equation}
\sqrt{-g}\left[L_{f}+
\frac{\lambda_{f}^{2}}{m_{\phi}^{2}}(\bar{\psi}\psi)^{2}\right]
+\frac{\lambda_{f}\lambda_{top}}{m_{\phi}^{2}}\Omega\bar{\psi}\psi
=0 \label{2-1-Dirac}
\end{equation}

Varying the effective action (\ref{2-1-L2}) with respect to
$g^{\mu\nu}$ and using Eq.(\ref{2-1-Dirac}) we obtain the
following fermion contribution to the energy-momentum tensor
proportional to $g_{\mu\nu}$:
\begin{equation}
\Delta T_{\mu\nu}^{(ferm)}= g_{\mu\nu}
\left[\frac{\lambda_{f}\lambda_{top}}{m_{\phi}^{2}}
\frac{\Omega}{\sqrt{-g}}\bar{\psi}\psi
+\frac{\lambda_{f}^{2}}{2m_{\phi}^{2}}(\bar{\psi}\psi)^{2}\right]
\label{2-1-contr}
\end{equation}

\subsection {\it A model with fermion selfinteraction}

Another model of interest is that of a fermion with a
selfinteraction in curved space-time. The fermion part of the
action we choose in the form
\begin{equation}
S=\int \sqrt{-g}d^{4}x\left[L_{f}+\lambda
(\bar{\psi}\psi)^l\right] \label{2-2-action}
\end{equation}
where $L_{f}$ is
 the free fermion Lagrangian density in the Einstein-Cartan
 space-time; $l$ is a dimensionless constant different from one;
 $\lambda$ is a coupling of the selfinteraction.

 It follows from the fermion  equation of motion that
\begin{equation}
L_{f}=-l\lambda (\bar{\psi}\psi)^l \label{2-2-Dirac}
\end{equation}
Varying the  action (\ref{2-2-action}) with respect to
$g^{\mu\nu}$ and using Eq.(\ref{2-2-Dirac}) we obtain the
following fermion contribution to the energy-momentum tensor
proportional to $g_{\mu\nu}$:
\begin{equation}
\Delta T_{\mu\nu}^{(ferm)}= (l-1)\lambda (\bar{\psi}\psi)^{l}
g_{\mu\nu} \label{2-2-Tmunu}
\end{equation}

The terms proportional to $g_{\mu\nu}$ in the above examples can
be regarded as possible candidates for fermion contributions to
the dark energy. However for this effect to be realized as a
genuine dark energy in the present day universe one needs that the
coefficient of the $g_{\mu\nu}$ term to be approximately a
space-time constant. But there are no reasons why these
coefficients in Eqs.(\ref{2-1-contr}) or (\ref{2-2-Tmunu}) should
be constant even approximately. As we will see in TMT this goal
can be achieved.

\section{Main Ideas of the Two Measures Theory}

TMT is a generally coordinate invariant theory where the action
has the form\cite{GK1}-\cite{GK5}
\begin{equation}
    S = \int L_{1}\Phi d^{4}x +\int L_{2}\sqrt{-g}d^{4}x
\label{S}
\end{equation}
 including two Lagrangians $ L_{1}$ and $L_{2}$ and two
measures of integration: the usual one $\sqrt{-g}$ and the new one
$\Phi$. The latter may be built by means   either of  four scalar
fields $\varphi_{a}$ ($a=1,2,3,4$) or of a totally antisymmetric
three index field $A_{\alpha\beta\gamma}$.

\subsection{Measure $\Phi$ built of  four scalar
fields $\varphi_{a}$ }

The measure  $\Phi$ being  a scalar density and a total derivative
may be defined as follows\footnote{As it was shown in the recent
paper\cite{Giddings}, the incorporation of four scalar fields
$\varphi_a$  together with the scalar density $\Phi$,
Eq.(\ref{Phi}), is  a possible way to define local observables in
the local quantum field theory approach to quantum gravity.}:
\begin{equation}
\Phi
=\varepsilon^{\mu\nu\alpha\beta}\varepsilon_{abcd}\partial_{\mu}\varphi_{a}
\partial_{\nu}\varphi_{b}\partial_{\alpha}\varphi_{c}
\partial_{\beta}\varphi_{d}.
\label{Phi}
\end{equation}
To provide parity conservation one can choose for example one of
$\varphi_{a}$'s to be a pseudoscalar. A shift of $L_{1}$ by a
constant, $L_{1}\rightarrow L_{1}+const.$, has no effect on the
equations of motion. Similar shift of $L_{2}$ would lead to the
change of the constant part of the Lagrangian coupled to the
volume element $\sqrt{-g}d^{4}x $. In  standard GR, this constant
term is the cosmological constant. However in TMT the relation
between the constant
 term of $L_{2}$ and the physical cosmological constant is very non
trivial.

In addition to the above idea concerning the general structure of
the action in TMT, there are only two basic assumptions:
\\
(1) The Lagrangian densities $ L_{1}$ and $L_{2}$ may be functions
of the matter fields,  the metric, the connection (or
spin-connection )
 but not of the "measure fields" $\varphi_{a}$. In such a case, i.e. when the
measure fields $\varphi_{a}$ enter in the theory only via the
measure $\Phi$,
  the action (\ref{S}) has
the infinite dimensional symmetry\cite{GK3}:
$\varphi_{a}\rightarrow\varphi_{a}+f_{a}(L_{1})$, where
$f_{a}(L_{1})$ is an arbitrary function of  $L_{1}$.
\\
(2) One should proceed in the first order formalism where all
fields, including metric and connections ( or vierbeins
$e^{\mu}_{a}$ and  spin-connection $\omega_{\mu}^{ab}$ in the
presence of fermions) as well as the measure fields $\varphi_{a}$
are independent dynamical variables. All the relations between
them follow from equations of motion. The independence of the
metric and the connection in the action means that we proceed in
the first order formalism and the relation between connection and
metric is not necessarily according to Riemannian geometry.

Varying the measure fields $\varphi_{a}$, we get
\begin{equation}
B^{\mu}_{a}\partial_{\mu}L_{1}=0 \label{var-phi}
\end{equation}
where
\begin{equation}
B^{\mu}_{a}=\varepsilon^{\mu\nu\alpha\beta}\varepsilon_{abcd}
\partial_{\nu}\varphi_{b}\partial_{\alpha}\varphi_{c}
\partial_{\beta}\varphi_{d}.
\label{Ama}
\end{equation}
Since $Det (B^{\mu}_{a}) = \frac{4^{-4}}{4!}\Phi^{3}$ it follows
that if $\Phi\neq 0$,
\begin{equation}
 L_{1}=sM^{4} =const
\label{varphi}
\end{equation}
where $s=\pm 1$ and $M$ is a constant of integration with the
dimension of mass.

\subsection{Measure $\Phi$ built of a totally antisymmetric
three index potential}

The measure $\Phi$ may be constructed also\cite{GK3} from a
totally antisymmetric three index potential
$A_{\alpha\beta\gamma}$:
\begin{equation}
\Phi
=\varepsilon^{\mu\nu\alpha\beta}\partial_{\mu}A_{\nu\alpha\beta}.
\label{Aabg}
\end{equation}
To provide parity conservation we must here choose
$A_{\alpha\beta\gamma}$ to have negative parity. The shift
symmetry $L_{1}\rightarrow L_{1}+const$ as well as two basic
assumptions similar to basic assumptions (1) and (2) of Sec.3.1
hold here too. However, the infinite dimensional symmetry
$\varphi_{a}\rightarrow\varphi_{a}+f_{a}(L_{1})$ is replaced now
by $A_{\alpha\beta\gamma}\rightarrow
A_{\alpha\beta\gamma}+\varepsilon_{\mu\alpha\beta\gamma}
f^{\mu}(L_{1})$ where $f^{\mu}(L_{1})$ are four arbitrary
functions of $L_{1}$ and $\varepsilon_{\mu\alpha\beta\gamma}$ is
numerically the same as $\varepsilon^{\mu\alpha\beta\gamma}$.

Variation of $A_{\alpha\beta\gamma}$ yields now
\begin{equation}
\varepsilon^{\mu\nu\alpha\beta}\partial_{\mu}L_{1}=0, \label{AdL1}
\end{equation}
that implies Eq.(\ref{varphi}) without the  condition $\Phi\neq 0$
needed in the model with four scalar fields $\varphi_{a}$.

\subsection{Generic features of TMT}

First of all one should notice
 {\it the very important differences of
TMT from scalar-tensor theories with nonminimal coupling}:

1. In general, the Lagrangian density $L_{1}$ (coupled to the
measure $\Phi$) may contain not only the scalar curvature term (or
more general gravity term) but also all possible matter fields
terms. This means that TMT {\it modifies in general both the
gravitational sector  and the matter sector};

2. If the field $\Phi$ were the fundamental (non composite) one
then the variation of $\Phi$ would result in the equation
$L_{1}=0$ instead of (\ref{varphi}), and therefore the
dimensionful parameter $M$ would not appear.

Applying the first order formalism one can show (see for example
Ref.\cite{GK3})  that the resulting relation between metric and
connection, as well as fermion and scalar fields equations contain
 the gradient of the ratio of the two measures
\begin{equation}
\zeta \equiv\frac{\Phi}{\sqrt{-g}} \label{zeta}
\end{equation}
which is a scalar field. It turns out that at least at the
classical level, the measure fields $\varphi_{a}$ (in the model of
Sec.3.1) or $A_{\alpha\beta\gamma}$ (in the model of Sec.3.2)
affect the theory only through the scalar field $\zeta$.

The consistency condition of equations of motion has the form of
algebraic constraint which determines $\zeta (x)$ as a function of
matter fields. The surprising feature of the theory is
 that neither Newton constant nor curvature appear in this constraint
which means that the {\it geometrical scalar field} $\zeta (x)$
{\it is determined by the matter fields configuration}  locally
and straightforward (that is without gravitational interaction).

By an appropriate change of the dynamical variables which includes
a conformal transformation of the metric, one can formulate the
theory in a Riemannian (or Riemann-Cartan) space-time. The
corresponding conformal frame we call "the Einstein frame". The
big advantage of TMT is that in the very wide class of models,
{\it the gravity and all matter fields equations of motion take
canonical GR form in the Einstein frame}.
 All the novelty of TMT in the Einstein frame as compared
with the standard GR is revealed only
 in an unusual structure of the scalar fields
effective potential, masses of fermions  and their interactions
with scalar fields as well as in the unusual structure of fermion
contributions to the energy-momentum tensor: they appear to be
$\zeta$ dependent. This is why the scalar field $\zeta (x)$
determined by the constraint, has a key role in these effects.

\section{Scale and Gauge Invariant Model}

To simplify the presentation of the main results we will study
here a simplified gauge model which is Abelian, does not include
the Higgs fields and quarks and chiral properties  of fermions are
ignored. In this simplified model the mass terms for Dirac spinors
are included by hand from the very beginning into the Lagrangians
$L_{1}$ and $L_{2}$.  The matter content of our model includes the
scalar field $\phi$, two so-called primordial fermion fields (the
primordial neutrino $N$ and the  primordial electron $E$) and
electromagnetic field $A_{\mu}$. The latter is included in order
to show the reasons why the gauge fields dynamics is canonical.
Generalization to non-Abelian gauge models including also quarks,
Higgs fields, the Higgs mechanism of mass generation  and taking
into account chiral properties  of fermions is straightforward,
see Ref.\cite{GK5}. The presence of the scalar field $\phi$ allows
to realize a spontaneously broken global scale invariance\cite{G1}
which includes the shift transformation of $\phi$. We will see
that in the cosmological context $\phi$ contributes to the present
dark energy.

We allow in both $L_{1}$ and $L_{2}$ all the usual terms
considered in standard field theory models in curved space-time.
Keeping the general structure of Eq.(\ref{S}) it is convenient to
represent the action in the following form:
\begin{eqnarray}
&S&=\int d^{4}x e^{\alpha\phi /M_{p}}(\Phi +b\sqrt{-g})
\left[-\frac{1}{\kappa}R(\omega ,e) +
\frac{1}{2}g^{\mu\nu}\phi_{,\mu}\phi_{,\nu}\right]
\nonumber\\
&& -\int d^{4}x e^{2\alpha\phi /M_{p}}[\Phi V_{1} +\sqrt{-g}V_{2}]
-\int
d^{4}x\sqrt{-g}\frac{1}{4}g^{\alpha\beta}g^{\mu\nu}F_{\alpha\mu}F_{\beta\nu}
\nonumber\\
&& +\int d^{4}x e^{\alpha\phi /M_{p}}(\Phi +k\sqrt{-g})
\frac{i}{2}\sum_{i}\overline{\Psi}_{i}
\left(\gamma^{a}e_{a}^{\mu}\overrightarrow{\nabla}^{(i)}_{\mu}-
\overleftarrow{\nabla}^{(i)}_{\mu}\gamma^{a}e_{a}^{\mu}\right)\Psi_{i}
\nonumber\\
     &&-\int d^{4}xe^{\frac{3}{2}\alpha\phi /M_{p}}
\left[(\Phi +h_{E}\sqrt{-g})\mu_{E}\overline{E}E +(\Phi
+h_{N}\sqrt{-g})\mu_{N}\overline{N}N \right] \label{totaction}
\end{eqnarray}
where $\Psi_{i}$ ($i=N,E$) is the general notation for the
primordial fermion fields $N$ and $E$;
$F_{\alpha\beta}=\partial_{\alpha}A_{\beta}-\partial_{\beta}A_{\alpha}$;
$\mu_{N}$ and $\mu_{E}$ are the mass parameters;
$\overrightarrow{\nabla}^{(N)}_{\mu}=\vec{\partial}+
\frac{1}{2}\omega_{\mu}^{cd}\sigma_{cd}$,
$\overrightarrow{\nabla}^{(E)}_{\mu}=\vec{\partial}+
\frac{1}{2}\omega_{\mu}^{cd}\sigma_{cd}+ieA_{\mu}$;
 $R(\omega ,V)
=e^{a\mu}e^{b\nu}R_{\mu\nu ab}(\omega)$ is the scalar curvature;
$e_{a}^{\mu}$ and $\omega_{\mu}^{ab}$ are the vierbein and
spin-connection; $g^{\mu\nu}=e^{\mu}_{a}e^{\nu}_{b}\eta^{ab}$ and
$R_{\mu\nu ab}(\omega)=\partial _{mu}\omega_{\nu
ab}+\omega^{c}_{\mu a}\omega_{\nu cb}-(\mu \leftrightarrow\nu)$.
$V_{1}$ and $V_2$ are constants with the dimensionality
$(mass)^4$. When Higgs field is included into the model then
$V_{1}$ and $V_2$ turn into functions of the Higgs field. As we
will see later , in the Einstein frame $V_{1}$, $V_2$  and
$e^{\alpha\phi /M_{p}}$ enter in the effective potential of the
scalar sector. Constants $b, k, h_{N}, h_{E}$  are non specified
dimensionless real parameters and we will only assume that they
 have close orders of magnitude
\begin{equation}
b\sim k\sim h_{N}\sim h_{E}; \label{sim-parameters}
\end{equation}
 $\alpha$ is a real parameter
which we take to be positive.

The action (\ref{totaction}) is invariant under the global scale
transformations:
\begin{eqnarray}
    &&e_{\mu}^{a}\rightarrow e^{\theta /2}e_{\mu}^{a}, \quad
\omega^{\mu}_{ab}\rightarrow \omega^{\mu}_{ab}, \quad
\varphi_{a}\rightarrow \lambda_{ab}\varphi_{b}\quad
A_{\alpha}\rightarrow A_{\alpha}, \nonumber
\\
 &&\phi\rightarrow
\phi-\frac{M_{p}}{\alpha}\theta ,\quad \Psi_{i}\rightarrow
e^{-\theta /4}\Psi_{i}, \quad \overline{\Psi}_{i}\rightarrow
e^{-\theta /4} \overline{\Psi}_{i}, \nonumber
\\
 &&where \quad \theta =const, \quad \lambda_{ab}=const \quad and
 \quad \det(\lambda_{ab})=e^{2\theta},
\label{stferm}
\end{eqnarray}
when the measure fields $\varphi_{a}$ are used for definition of
the measure $\Phi$, as in Sec.3.1. If the definition (\ref{Aabg})
of Sec.3.2 is used then the scale transformation of the totally
antisymmetric three index potential $A_{\alpha\beta\gamma}$ should
be: $A_{\alpha\beta\gamma}\rightarrow
e^{2\theta}A_{\alpha\beta\gamma}$.

The choice of the action (\ref{totaction}) needs two observations:

1. We have chosen the kinetic term of $A_{\mu}$ in the conformal
invariant form which is possible  if it is coupled only to the
measure $\sqrt{-g}$. Introducing the coupling of this term to the
measure $\Phi$ would lead to the nonlinear field strength
dependence  in the $A_{\mu}$ equation of motion. One of the
possible consequences of this  may be  non positivity of the
energy density. Another consequence is a possibility of certain
unorthodox effects, like space-time variations of the effective
fine structure constant. This subject deserves a special study but
it is out of the purposes of this paper where the model is studied
setting $F_{\mu\nu}\equiv 0$ in the solutions.

2. With the aim to simplify the analysis of the results of the
model (containing many free parameters) we have chosen the
coefficient $b$ in front of $\sqrt{-g}$ in the first integral of
(\ref{totaction}) to be a common factor of the gravitational term
 $-\frac{1}{\kappa}R(\omega ,e)$ and of the kinetic term for the
  field $\phi$. The research of more general case has been started in
  Ref.\cite{GK4}. A more detailed study of interesting consequences of
  this possibility will be presented in a separate paper\cite{GK6}.

 We would like to stress that
except for the modification of the general structure of the action
based on the basic assumptions of TMT,
 {\it we do not introduce into the action}
(\ref{totaction}) {\it any exotic terms and  fields}. On the other
hand, in the framework of such an economic  model, except for the
above two notions, Eq.(\ref{totaction}) describes {\em the most
general action of TMT satisfying the formulated above symmetries}.

Variation of the measure fields ($\varphi_{a}$ in the model of
Sec.3.1 or $A_{\alpha\beta\gamma}$ in the model of Sec.3.2) yields
Eq.(\ref{varphi}) where $L_{1}$ is now defined, according to
Eq.(\ref{S}), as the part of the integrand of the action
(\ref{totaction}) coupled to the measure $\Phi$. The appearance of
the integration constant $sM^{4}$ in Eq.(\ref{varphi})
spontaneously breaks the global scale invariance (\ref{stferm}).
In what follows we choose $s=+1$.

Except for the $A_{\mu}$ equation, all other equations of motion
resulting from (\ref{totaction}) in the first order formalism
contain terms proportional to $\partial_{\mu}\zeta$ that makes the
space-time non-Riemannian and equations of motion - non canonical.
However, with the new set of variables ($\phi$ and $A_{\mu}$
remain unchanged)
\begin{eqnarray}
&&\tilde{e}_{a\mu}=e^{\frac{1}{2}\alpha\phi/M_{p}}(\zeta
+b)^{1/2}e_{a\mu}, \quad
\tilde{g}_{\mu\nu}=e^{\alpha\phi/M_{p}}(\zeta +b)g_{\mu\nu},
\nonumber\\
&&\Psi^{\prime}_{i}=e^{-\frac{1}{4}\alpha\phi/M_{p}} \frac{(\zeta
+k)^{1/2}}{(\zeta +b)^{3/4}}\Psi_{i} , \quad i=N,E \label{ctferm}
\end{eqnarray}
which we call the Einstein frame,
 the spin-connections become those of the
Einstein-Cartan space-time. Since $\tilde{e}_{a\mu}$,
$\tilde{g}_{\mu\nu}$, $N^{\prime}$ and $E^{\prime}$ are invariant
under the scale transformations (\ref{stferm}), spontaneous
breaking of the scale symmetry (by means of Eq.(\ref{varphi})) is
reduced in the new variables to the {\it spontaneous breaking of
the shift symmetry}
\begin{equation}
\phi\rightarrow\phi +const. \label{phiconst}
\end{equation}
Notice that the Goldstone theorem generically is not applicable in
this theory (see the second reference in Ref.\cite{G1})). The
reason is the following. In fact, the scale symmetry
(\ref{phiconst}) leads to a conserved dilatation current
$j^{\mu}$. However, for example in the spatially flat FRW universe
the spatial components of the current $j^{i}$ behave as
$j^{i}\propto M^4x^i$ as $|x^i|\rightarrow\infty$. Due to this
anomalous behavior at infinity, there is a flux of the current
leaking to infinity, which causes the non conservation of the
dilatation charge. The absence of the latter implies that one of
the conditions necessary for the Goldstone theorem is missing. The
non conservation of the dilatation charge is similar to the well
known effect of instantons in QCD where singular behavior in the
spatial infinity leads to the absence of the Goldstone boson
associated to the $U(1)$ symmetry.

 After the change of
variables (\ref{ctferm}) to the Einstein frame and some simple
algebra, the gravitational equations  take the standard GR form
\begin{equation}
G_{\mu\nu}(\tilde{g}_{\alpha\beta})=\frac{\kappa}{2}T_{\mu\nu}^{eff}
 \label{gef}
\end{equation}
where  $G_{\mu\nu}(\tilde{g}_{\alpha\beta})$ is the Einstein
tensor in the Riemannian space-time with the metric
$\tilde{g}_{\mu\nu}$; the energy-momentum tensor
$T_{\mu\nu}^{eff}$ is now
\begin{eqnarray}
T_{\mu\nu}^{eff}&=&\phi_{,\mu}\phi_{,\nu}-\frac{1}{2}
\tilde{g}_{\mu\nu}\tilde{g}^{\alpha\beta}\phi_{,\alpha}\phi_{,\beta}
+\tilde{g}_{\mu\nu}V_{eff}(\phi;\zeta)
\nonumber\\
&+&T_{\mu\nu}^{(em)}
+T_{\mu\nu}^{(ferm,can)}+T_{\mu\nu}^{(ferm,noncan)};
 \label{Tmn}
\end{eqnarray}
\begin{equation}
V_{eff}(\phi;\zeta)=
\frac{b\left(M^{4}e^{-2\alpha\phi/M_{p}}+V_{1}\right)
-V_{2}}{(\zeta +b)^{2}}; \label{Veff1}
\end{equation}
$T_{\mu\nu}^{(em)}$ is the canonical energy momentum tensor of the
electromagnetic field; $T_{\mu\nu}^{(ferm,can)}$ is the canonical
energy momentum tensor for (primordial) fermions $N^{\prime}$ and
$E^{\prime}$ in curved space-time (including also interaction of
$E^{\prime}$ with $A_{\mu}$). $T_{\mu\nu}^{(ferm,noncan)}$ is the
{\em noncanonical} contribution of the fermions into the energy
momentum tensor
\begin{equation}
 T_{\mu\nu}^{(ferm,noncan)}=-\tilde{g}_{\mu\nu}\Lambda_{dyn}^{(ferm)}
 \label{Tmn-noncan}
\end{equation}
where
\begin{equation}
\Lambda_{dyn}^{(ferm)}\equiv Z_{N}(\zeta)m_{N}(\zeta)
\overline{N^{\prime}}N^{\prime}+
Z_{E}(\zeta)m_{E}(\zeta)\overline{E^{\prime}}E^{\prime}
\label{Lambda-ferm}
\end{equation}
and $Z_{i}(\zeta)$ and $m_{i}(\zeta)$ ($i=N^{\prime},E^{\prime}$)
are respectively
\begin{equation}
Z_{i}(\zeta)\equiv \frac{(\zeta -\zeta^{(i)}_{1})(\zeta
-\zeta^{(i)}_{2})}{2(\zeta +k)(\zeta +h_{i})}, \qquad
m_{i}(\zeta)= \frac{\mu_{i}(\zeta +h_{i})}{(\zeta +k)(\zeta
+b)^{1/2}} \label{Zeta&m}
\end{equation}
where
\begin{equation}
\zeta_{1,2}^{(i)}=\frac{1}{2}\left[k-3h_{i}\pm\sqrt{(k-3h_{i})^{2}+
8b(k-h_{i}) -4kh_{i}}\,\right].
 \label{zeta12}
\end{equation}

The noncanonical contribution $T_{\mu\nu}^{(ferm,noncan)}$ of the
fermions into the energy momentum tensor has the transformation
properties of a cosmological constant term but it is proportional
to fermion densities
$\overline{\Psi}^{\prime}_{i}\Psi^{\prime}_{i}$ \,
($i=N^{\prime},E^{\prime}$). This is why we will refer to it as
"dynamical fermionic $\Lambda$ term". This fact is displayed
explicitly in Eqs.(\ref{Tmn-noncan}),(\ref{Lambda-ferm}) by
defining $\Lambda_{dyn}^{(ferm)}$.

$T_{\mu\nu}^{(ferm,noncan)}$ is somewhat similar to the
contributions $\Delta T_{\mu\nu}$ of the fermions into the energy
momentum tensor obtained in the non-TMT models of Sec.2. But there
are the following essential differences:
\begin{itemize}

\item The appearance of $T_{\mu\nu}^{(ferm,noncan)}$
in Eq.(\ref{Tmn}) is a direct result of the action
(\ref{totaction}) while the contribution $\Delta T_{\mu\nu}$ in
the first model of Sec.2 results from integrating out the slowly
varying scalar field; in the second model of Sec.2 we introduced
into the action the fermion self-interaction while our TMT action
(\ref{totaction}) has no such exotic terms.

\item The contribution $\Delta T_{\mu\nu}$ in the  models of Sec.2
may be a source of departures from GR. On the contrary, as we will
see in Sec.6, $\Lambda_{dyn}^{(ferm)}$ becomes negligible in
gravitational experiments with regular matter.

\item As we have noted in Sec.2, there are no reasons why the
coefficient of $g_{\mu\nu}$ in $\Delta T_{\mu\nu}$ in the models
of Sec.2 might be constant even approximately. But  we will show
in the context of a toy cosmological model in Sec.7 that a
neutrino contribution into $T_{\mu\nu}^{(ferm,noncan)}$ may  be a
part of the cosmological constant term at the very late universe.

\end{itemize}

The "dilaton" $\phi$ field equation  in the Einstein frame reads
\begin{equation}
\Box\phi -\frac{\alpha}{M_{p}(\zeta +b)}
\left[M^{4}e^{-2\alpha\phi/M_{p}}-\frac{(\zeta -b)V_{1}(\upsilon)
+2V_{2}(\upsilon)}{\zeta +b}\right]= -\frac{\alpha
}{M_{p}}\Lambda_{dyn}^{(ferm)}, \label{phief+ferm1}
\end{equation}
where $\Box\phi =(-\tilde{g})^{-1/2}\partial_{\mu}
(\sqrt{-\tilde{g}}\tilde{g}^{\mu\nu}\partial_{\nu}\phi)$.

One can show that equations  for the primordial fermions in terms
of the variables (\ref{ctferm})
 take the standard form
of fermionic equations for $N^{\prime}$ and $E^{\prime}$ in the
Einstein-Cartan space-time (see also Appendix A)  where the
standard electromagnetic interaction of $E^{\prime}$ is present
too. Note that the coupling of the fermions to the dilaton $\phi$
in the original action (which was only via exponents of $\phi$)
disappear in the Einstein frame. All the novelty
 as compared with the standard field theory
approach  consists of the form of the $\zeta$ depending "masses"
of the primordial fermions $N^{\prime}$, $E^{\prime}$, second
equation in Eq.(\ref{Zeta&m}). The  fermion parts of the effective
action in the Einstein frame is invariant under the global phase
transformations of any of fermion fields exactly as it was in the
original action. Therefore the conserved fermion $4$-currents
exist both in the original and in the Einstein frames. However in
the Einstein frame these currents are not only  covariantly
conserved but also have the canonical form without any $\phi$ and
$\zeta$ dependence.

The  electromagnetic field equations are canonical due to our
choice  for the appropriate term in the action (\ref{totaction})
to be conformally invariant.

The  scalar field $\zeta$ (see its definition, Eq.(\ref{zeta})),
is determined as a function of the scalar field $\phi$ and
$\overline{\Psi}^{\prime}_{i}\Psi^{\prime}_{i}$ \,
($i=N^{\prime},E^{\prime}$) by the following constraint
\begin{equation}
\frac{1}{(\zeta
+b)^{2}}\left\{(b-\zeta)\left[M^{4}e^{-2\alpha\phi/M_{p}}+
V_{1}(\upsilon)\right]-2V_{2}(\upsilon)\right\}=
\Lambda_{dyn}^{(ferm)} \label{constraint}
\end{equation}
The origin of the constraint is clear enough. There are two
equations containing the scalar curvature: the first one is
Eq.(\ref{varphi}) and the second one follows from gravitational
equations. The constraint is nothing but the consistency condition
of these two equations.

One should point out an unexpected and very important fact, namely
that the same function $\Lambda_{dyn}^{(ferm)}$,
Eq.(\ref{Lambda-ferm}), emerges in
 three different places:

a) in the form of the noncanonical fermion contribution
 to the energy-momentum tensor, Eq.(\ref{Tmn-noncan});

b)  in the
 effective Yukawa coupling of the dilaton $\phi$ to fermions
 (see the right hand side of Eq.(\ref{phief+ferm1}));

c) as the right hand side of the constraint.

Note that the original action (\ref{totaction}) contains exponents
of the scalar field $\phi$ and in particular the coupling of
fermions with $\phi$ is realized in (\ref{totaction}) only through
the exponents of $\phi$. Nevertheless, except for the term
$M^{4}e^{-2\alpha\phi/M_{p}}$ originated by the scale symmetry
breaking, the equations of motion  in the Einstein frame do not
contain explicitly the exponents of $\phi$ . However, the
Yukawa-type coupling of the fermions to $\phi$ emerges in the
Einstein frame, see the r.h.s. of Eq.(\ref{phief+ferm1}).

It is interesting that non-explicit dependence on the exponent of
$\phi$  in the equations of motion is actually present after
solving the constraint (\ref{constraint}) for $\zeta$. However
this dependence is again in the form of
$M^{4}e^{-2\alpha\phi/M_{p}}$. Thus the exponential
$\phi$-dependence in the equations of motion results only from the
scale symmetry breaking. Recall that in the Einstein frame the
scale symmetry transformations (\ref{stferm}) are reduced to the
shift symmetry $\phi\rightarrow\phi + const$. Note finally that
applying the constraint (\ref{constraint}) to
Eq.(\ref{phief+ferm1}) one can reduce the latter to the form
\begin{equation}
\Box\phi-\frac{2\alpha\zeta}{(\zeta
+b)^{2}M_{p}}M^{4}e^{-2\alpha\phi/M_{p}} =0, \label{phi-constr}
\end{equation}
where $\zeta$  is a solution of the constraint (\ref{constraint}).
 This result is true both in the presence of
fermions and in their absence.

Generically, the constraint (\ref{constraint}) determines $\zeta$
as a complicated function of $\phi$,
$\overline{N^{\prime}}N^{\prime}$ and
$\overline{E^{\prime}}E^{\prime}$. Substituting the appropriate
solution for $\zeta$ into the equations of motion one can conclude
that in general there is no sense, for example, to regard
$V_{eff}(\phi,\upsilon ;\zeta)$, Eq.(\ref{Veff1}), as the
effective potential for the scalar field $\phi$ because it depends
in a very nontrivial way on $\overline{N^{\prime}}N^{\prime}$ and
$\overline{E^{\prime}}E^{\prime}$ as well. For the same reason,
the fermion mass and $\Lambda_{dyn}^{(ferm)}$ describe in general
self-interactions of the primordial fermions depending also on the
scalar field $\phi$. Therefore it is impossible, in general, to
separate the terms of $T_{\mu\nu}$ describing the scalar field
$\phi$ effective potential from the fermion contributions. Such
mixing of the scalar field $\phi$ associated with dark energy, on
the one hand, and fermionic matter, on the other hand, gives rise
to a rather complicated system of equations when trying to apply
the theory to general situations that could appear in astrophysics
and cosmology. Notice that in such a case, the quantum theory of
fermion fields may be non perturbative: inserting solution for
$\zeta$ into the effective fermion "mass", Eq.(\ref{Zeta&m}), it
is easy to see that the "free" primordial fermion equation appears
to be nonlinear in general. Considerable simplification of the
situation occurs if for some reasons $\zeta$  appears to be
constant or almost constant. Fortunately this is exactly what
happens in many physically interesting situations.

\section{Dark Energy in the Absence of Massive Fermions}

In the absence of massive fermions the constraint
(\ref{constraint}) determines $\zeta$ as the function of $\phi$
alone:
\begin{equation}
\zeta =\zeta_{0}(\phi)\equiv b-\frac{2V_{2}}
{V_{1}+M^{4}e^{-2\alpha\phi/M_{p}}}, \label{zeta-without-ferm}
 \end{equation}
Note that the electromagnetic field does not enter in the
constraint (\ref{constraint}) and therefore the presence of the
electromagnetic field does not affect the value of $\zeta_{0}$.

 The effective potential of
the scalar field $\phi$ results then from Eq.(\ref{Veff1})
\begin{equation}
V_{eff}^{(0)}(\phi)\equiv
V_{eff}(\phi;\zeta_{0})|_{\overline{\psi^{\prime}}\psi^{\prime}=0}
=\frac{[V_{1}+M^{4}e^{-2\alpha\phi/M_{p}}]^{2}}
{4[b\left(V_{1}+M^{4}e^{-2\alpha\phi/M_{p}}\right)-V_{2}]}
\label{Veffvac}
\end{equation}
and the $\phi$-equation (\ref{phief+ferm1}) is reduced to
\begin{equation}
\Box\phi +V^{(0)\prime}_{eff}(\phi)=0,
\label{eq-phief-without-ferm}
\end{equation}
where prime sets derivative with respect to $\phi$.

 The mechanism of the appearance of the effective potential (\ref{Veffvac})
 is very interesting and exhibits the main features of our TMT
 model\footnote{The particular case of this model
with $b=0$ and $V_{2}<0$ was studied in Ref.\cite{G1}. The
application of the TMT model with explicitly broken global scale
symmetry to the quintessential inflation scenario was discussed in
Ref.\cite{K}. A particular case of the model (\ref{totaction})
without explicit potentials, i.e. $V_{1}=V_{2}=0$, has been
studied in Ref.\cite{GK4}; see also Appendix of the present
paper.}.
 In fact, the reasons of the transformation of the
 prepotentials $V_{1}e^{2\alpha\phi/M_{p}}$
 and $V_{2}e^{2\alpha\phi/M_{p}}$, coming in the original action
 (\ref{totaction}),  into the effective potential (\ref{Veffvac})
 are the following:

a) Transformation to the Einstein frame;

b)  Spontaneous breakdown of the global scale symmetry which in
the Einstein frame is reduced to the spontaneously broken shift
symmetry (\ref{phiconst});

c) The constraint which in the absence of fermions case takes the
form (\ref{zeta-without-ferm}).

It is easy to see that the gravitational equations (\ref{gef}) in
the absence of fermions case become the standard Einstein
equations of the model where the electromagnetic field and the
minimally coupled scalar field $\phi$ with the potential
$V_{eff}^{(0)}(\phi)$ are the sources of gravity.

The structure of the potential (\ref{Veffvac}) allows to construct
a model\cite{G1},\cite{GK3} where zero vacuum energy is achieved
without fine tuning when $V_{1}+M^{4}e^{-2\alpha\phi/M_{p}}=0$.
This and many other aspects of the dynamics of the scalar sector
(including Higgs fields) will be studied in detail in a separate
publication\cite{GK6}.

In what follows we will assume
\begin{equation}
V_{1}>0 \quad and \quad b>0. \label{param-vac}
\end{equation}
 Applying this model to the cosmology
of the late time universe and assuming that the scalar field
$\phi\rightarrow\infty$ as $t\rightarrow\infty$, we see that the
evolution of the late time universe  is governed by the dark
energy density
\begin{equation}
\rho^{(0)}_{d.e}=\frac{1}{2}\dot{\phi}^{2}+\Lambda^{(0)}+V^{(0)}_{q-like}(\phi).
\label{rho-without-ferm}
\end{equation}
where $\Lambda^{(0)}$ is the cosmological constant
\begin{equation}
\Lambda^{(0)} =\frac{V_{1}^{2}} {4[bV_{1}-V_{2}]}
\label{lambda-without-ferm}
\end{equation}
and $V^{(0)}_{q-like}(\phi)$ is the quintessence-like scalar field
 potential
\begin{equation}
V^{(0)}_{q-like}(\phi) =\frac{V_{1}(bV_{1}-2V_{2})+
(bV_{1}-V_{2})M^{4}e^{-2\alpha\phi/M_{p}}}
{4[bV_{1}-V_{2}][b(V_{1}+
M^{4}e^{-2\alpha\phi/M_{p}})-V_{2}]}\cdot \,
M^{4}e^{-2\alpha\phi/M_{p}}. \label{V-quint-without-ferm}
\end{equation}

The cosmological constant $\Lambda^{(0)}$ is the asymptotic value
( as $t\rightarrow\infty$) of $\rho^{(0)}_{d.e}$ for the FRW
universe in the model where massive fermions absent (the reason we
emphasize this here will be clear in Sec.7). $\Lambda^{(0)}$ is
positive provided $bV_{1}>V_{2}$. The potential decreases to
$\Lambda^{(0)}$ monotonically if
\begin{equation}
bV_{1}>2V_{2} \label{bv1>2v2}
\end{equation}
that will be assumed in what follows.

There are two ways to provide the observable order of magnitude of
the present day vacuum energy density by an appropriate choice of
the parameters of the theory in the framework of the condition
(\ref{bv1>2v2}) but without supposition of an extreme smallness of
$V_1$ and/or $V_2$:

1. If $V_2<0$ and $bV_{1}<|V_{2}|$ then $\Lambda^{(0)}\approx
\frac{V_1^2}{4|V_2|}$. In this case there is no need for $V_{1}$
and $V_{2}$ to be small: it is enough that the dimensionless
quantity
 $V_{1}/ |V_{2}|\ll 1 $.  This possibility is a kind of  {\it seesaw}
 mechanism
(see Refs.\cite{G1},\cite{see}). For instance, if $V_{1}$ has the
  scale of electroweak symmetry breaking $V_{1}\sim
(10^{3}GeV)^{4}$ and $V_{2}$ has   the Planck scale $|V_{2}| \sim
(10^{18}GeV)^{4}$ then $\Lambda^{(0)}\sim (10^{-3}eV)^{4}$.

2. If $V_2>0$ or alternatively $V_2<0$ and $bV_{1}>|V_{2}|$ then
$\Lambda^{(0)}\approx \frac{V_1}{4b}$. Hence the second
possibility is to choose the {\it dimensionless} parameter $b>0$
to be a huge number.   In this case the order of magnitudes of
$V_{1}$ and $V_{2}$ could be either as in the above case (i) or to
be  not too much different (or even of the same order). For
example, if $V_{1}\sim (10^{3}GeV)^{4}$ then for getting
$\Lambda^{(0)}\sim (10^{-3}eV)^{4}$ one should assume that $b\sim
10^{60}$. Note that $b$ is the ratio of the coupling constants of
the scalar curvature to the measures $\sqrt{-g}$ and $\Phi$ in the
fundamental action of the theory (\ref{totaction}).

\section{General Relativity and Regular Fermions}

\subsection{Reproducing Einstein Equations and Regular Fermions}

In Sec.5 we have seen that in the absence of fermions case the
gravitational equations (\ref{gef}) coincide with the Einstein
equations. Analyzing Eqs.(\ref{gef})-(\ref{zeta12}) in more
general cases it is easy to see that Eqs.(\ref{gef}) and
(\ref{Tmn}) are reduced to the Einstein equations in the
appropriate field theory model (i.e. when the scalar field,
electromagnetic field and massive fermions are sources of gravity)
if $\zeta$ is constant and
\begin{equation}
\Lambda_{dyn}^{ferm}=0 \qquad or\quad at\quad least\qquad
|T_{\mu\nu}^{(ferm,noncan)}|\ll |T_{\mu\nu}^{(ferm,can)}|.
\label{noncan-ll-can}
\end{equation}

According to Eqs.(\ref{Tmn-noncan})-(\ref{zeta12}), in the case
when a single massive fermion is a source of gravity, the
condition (\ref{noncan-ll-can}) is realized  if
\begin{equation}
Z_{i}(\zeta)\approx 0 \quad \Longrightarrow \quad
\zeta=\zeta_{1}^{i} \quad or \quad \zeta=\zeta_{2}^{i}, \quad
i=N^{\prime},E^{\prime}, \label{Z-0}
\end{equation}
where $\zeta_{1,2}^{i}$ are defined in Eqs.(\ref{zeta12}).

Recall that existence of a noncanonical contribution to the
energy-momentum tensor (\ref{Tmn-noncan}), along with the $\zeta$
dependence of the fermion mass (\ref{Zeta&m}) discovered in Sec.4,
displays  the fact that generically primordial fermion is very
much different from the regular one  (see definitions in item (ii)
of the Introduction section). To answer the question what are the
characteristic features of the regular massive fermion we have to
take into account the undisputed fact that the classical tests of
GR deal only with regular fermionic matter. Hence {\it we should
identify the regular fermions with states of the primordial
fermions satisfying the  condition} (\ref{Z-0}).

\subsection{Meaning of the Constraint and Regular Fermions}

We are going  now  to understand the meaning of the constraint
(\ref{constraint}). We start from the detailed analysis of two
limiting cases

(1) in space-time regions without  fermions;

(2) in space-time regions occupied by regular fermions;\\
and afterwards we will be able to formulate the meaning of the
constraint in general case. Recall that the main goal of this
paper realized in Sec.7 is to demonstrate a possibility of exotic
states where fermion affects the dark energy, and this result is
also based essentially on Eq.(\ref{constraint}).

We will proceed keeping in mind that the parameters $V_1$, $V_2$
and/or $b$ are chosen appropriately, as in the discussion in items
(i) or (ii) at the end of Sec.5 to provide a desirable order of
magnitude of the cosmological constant $\Lambda^{(0)}$ in the
absence of fermions case, Eq.(\ref{lambda-without-ferm}).

 It is convenient to
divide the analysis into a few steps:

1. It follows from  the condition (\ref{bv1>2v2}) that
$\zeta_{0}(\phi)$ (determined by Eq.(\ref{zeta-without-ferm})) has
the same order of magnitude as the parameter $b$.

2. Recall that $V_{eff}^{(0)}(\phi)$ having the order of magnitude
typical for the dark energy density (in the absence of fermions
case) is obtained from $V_{eff}(\phi;\zeta)$, Eq.(\ref{Veff1}), as
$\zeta =\zeta_{0}(\phi)$. Therefore with the help of the item (i)
we conclude that  each time when $\zeta$ has  the order of
magnitude close to that of the parameter $b$ (and if no special
tuning is assumed) $V_{eff}(\phi;\zeta)$, Eq.(\ref{Veff1}), has
the order of magnitude close to that of the dark energy density
(in the absence of fermions case).

3. It is easy to see that  {\it each time when} $\zeta$ {\it has
the order of magnitude close to that of the parameter} $b$ (if no
special tuning is assumed and in particular $\zeta\neq
\zeta_{0}(\phi)$), the left hand side (l.h.s.) of the constraint
(\ref{constraint}) has the order of magnitude close to that of
$V_{eff}(\phi;\zeta)$, Eq.(\ref{Veff1}), i.e. {\it the l.h.s. of
the constraint has the order of magnitude close to that of the
dark energy density in the absence of fermions case}.

4. Let us now turn  to the right hand side (r.h.s.) of the
constraint (\ref{constraint}) in the presence of a single massive
primordial fermion. It contains factor
$m_{i}(\zeta)\overline{\Psi}^{\prime}_{i}\Psi^{\prime}_{i}$, \,
($i=N^{\prime},E^{\prime}$) which have typical order of magnitude
of the fermion canonical energy density $T_{00}^{(ferm,can)}$.
 If the
primordial fermion is in a state of a regular fermion then
according to the conclusion made at the end of the previous
subsection,  in the space-time region where the fermion is
localized, the scalar $\zeta$ must be $\zeta=\zeta_{1}^{i}$ (or
$\zeta=\zeta_{2}^{i}$).  Therefore in the space-time region
occupied by a single regular fermion, the r.h.s. of
(\ref{constraint}) is
\begin{equation}
\Lambda_{dyn}^{(ferm)}|_{regular}\equiv
Z_{i}(\zeta_{1,2}^{i})m_{i}(\zeta_{1,2}^{i})
\left(\overline{\Psi^{\prime}_{i}}\Psi^{\prime}_{i}\right)_{regular}
\label{rhs-reg}
\end{equation}
Due to our assumption (\ref{sim-parameters}) it follows from the
definitions (\ref{zeta12}) that both $\zeta_{1}^{i}$ and
$\zeta_{2}^{i}$ have the order of magnitude close to that of $b$.
Hence in the space-time region occupied by a single regular
fermion, the l.h.s. of (\ref{constraint}) has the order of
magnitude close to that of the dark energy density in the fermion
vacuum. It is evident that in normal particle physics conditions,
that is when the energy density of a single fermion $\sim
m_{i}(\zeta)\overline{\Psi}^{\prime}_{i}\Psi^{\prime}_{i}$
 is tens of orders of magnitude larger than the
fermion vacuum energy density, the balance dictated by the
constraint is satisfied in the present day universe just due to
the condition (\ref{Z-0}).

5. In more general cases, i.e. when primordial fermion is in a
state different from the regular one, the meaning of the
constraint is similar: {\it the balance between the scalar dark
energy contribution  to the l.h.s. of the constraint and  the
 fermionic contribution to
the r.h.s. of the constraint} is realized due to the factors
$Z_{i}(\zeta)$. In other words, {\it the constraint describes the
local balance between the fermion energy density and the scalar
dark energy density} in the space-time region where the  wave
function of the primordial fermion is not equal to zero; {\it by
means of this balance the constraint determines the scalar}
$\zeta(x)$.  Note also that due to this balance, {\it the degree
of localization of the fermion and values of $\zeta(x)$ may be
strongly interconnected}. As we will see in Sec.7, this feature of
fermions in TMT plays {\it a key role in the mechanism providing a
possibility for a primordial fermion to be either in the state of
a regular fermion or in exotic states}.

One can suggest the following two alternative approaches to the
question of how a primordial fermion can be realized as a regular
one:

{\it The first approach} discussed in Refs.\cite{GK4}, \cite{GK5},
is based on the idea of the "maximal economy". We start from one
primordial fermion field for each type of fermions: one neutral
primordial lepton field $N$, one charged primordial lepton field
$E$ and similar for quarks. In other words we start from one
 generation of fermions (note that the gauge symmetry,
 for example $SU(2)\times U(1)$, may be imposed in a usual way).
Splitting of the primordial fermions into families occurs only in
normal particle physics conditions, i.e. when the fermion energy
density is huge in comparison with the vacuum energy density. One
of the possibilities for this to be realized is the above
mentioned condition $Z_{i}(\zeta)\approx 0$. The appropriate two
constant solutions for $\zeta$, i.e. $\zeta=\zeta_{1,2}^{i}$,
correspond to two different states of the primordial fermions with
{\it different constant masses} determined by the second equation
in (\ref{Zeta&m}) where we have to substitute $\zeta_{1,2}^{i}$
instead of $\zeta$. So, in the normal particle physics conditions,
the scalar $\zeta$ plays the role of an additional degree of
freedom determining different mass eigenstates  of the primordial
fermions which we want to identify with different fermion
generations. Note that the classical tests of GR deal in fact with
matter built of the fermions of the first generation (may be with
a small touch of the second generation). This is why one can
identify the states of the primordial fermions realized as $\zeta
=\zeta_{1,2}^{i}$ with the first two generations of the regular
fermions. For example, if the free primordial electron $E$ is in
the state with $\zeta =\zeta_{1}^{(E)}$ (or $\zeta
=\zeta_{2}^{(E)}$), it is detected as the regular electron $e$ (or
muon $\mu$) and similar the primordial neutrino $N$ splits into
the regular electron and muon neutrinos with masses respectively:
\begin{equation}
 m_{e(\mu)}=
\frac{\mu_{E}(\zeta_{1(2)}^{(E)} +h_{E})}{(\zeta_{1(2)}^{(E)}
+k)(\zeta_{1(2)}^{(E)} +b)^{1/2}};\quad m_{\nu_{e}(\nu_{\mu})}=
\frac{\mu_{N}(\zeta_{1(2)}^{(N)} +h_{N})}{(\zeta_{1(2)}^{(N)}
+k)(\zeta_{1(2)}^{(N)} +b)^{1/2}}\label{m-12}
\end{equation}
It turns out that  there is only one more additional possibility
to satisfy  the constraint (\ref{constraint}) when primordial
fermion is in the normal particle physics conditions. This is the
solution $\zeta^{i}=\zeta_{3}^{i}\approx -b$ which one can
associate with the third generation of fermions (for details see
Refs.\cite{GK4}, \cite{GK5}). It is interesting that in contrast
to the first two generations, the third generation defined by this
way, may have gravitational interaction with unusual features
since the condition (\ref{noncan-ll-can}) may not hold. The
described splitting of the primordial fermions into three
generations in the normal particle physics conditions is the
family replication mechanism proposed in Refs.\cite{GK4},
\cite{GK5}.

{\it The second approach} is based on the idea that the three
families of fermions of the standard model exist from the
beginning in the original action, i.e. not to use the family
replication mechanism for explanation of the observed three
generations of fermions. In this case again, exactly as it was in
the first approach, the primordial fermions turn into the regular
fermions only in the normal particle physics conditions. Now
however if we interpret the state of the primordial fermions with,
for example, $\zeta^{(i)} =\zeta_{1}^{(i)}$ as the observable
regular fermions, then some role should be assigned to the states
with $\zeta^{(i)} =\zeta_{2}^{(i)}$ and $\zeta^{(i)}
=\zeta_{3}^{(i)}$. By means of a choice of the parameters one can
try for example to provide very large masses of the regular
fermions with $\zeta^{(i)} =\zeta_{2}^{(i)}$ and $\zeta^{(i)}
=\zeta_{3}^{(i)}$ that might explain the reason why they are
unobservable so far. However these questions are beyond of the
goals of this paper and  together with many other aspects of
fermions in TMT will be studied in a separate publication.

\subsection{Resolution of the 5-th Force Problem for Regular Fermions}

Reproducing Einstein equations when the primordial fermions are in
the states of the regular fermions is not enough in order to
assert that GR is reproduced. The reason is that at the late
universe, as $\phi\gg M_{p}$, the scalar field $\phi$ effective
potential is very flat and therefore due to the Yukawa-type
coupling of massive fermions to $\phi$, (the r.h.s. of
Eq.(\ref{phief+ferm1})),  the long range scalar force appears to
be possible in general. The Yukawa coupling "constant" is
$\alpha\frac{m_{i}(\zeta)}{M_{p}}Z_{i}(\zeta)$. Applying our
analysis of the meaning of the constraint
 in Sec.6.2, it is easy to see that  for
regular fermions with $\zeta^{(i)} =\zeta_{1,2}^{(i)}$ the factor
$Z_{i}(\zeta)$ is of the order of the ratio of the vacuum energy
density to the regular fermion energy density. Thus we conclude
that the 5-th force is extremely suppressed for the fermionic
matter observable in classical tests of GR. It is very important
that this result is obtained automatically, without tuning of the
parameters and it takes place for both approaches to realization
of the regular fermions in TMT discussed in subsection 6.2.

\section{Nonrelativistic Neutrinos and Dark Energy}

In Secs.5 and 6 we have studied two opposite limiting cases: one
is realized if there are no  fermions at all; the second one
corresponds to the normal particle physics conditions. The latter
means that for a single fermion
$m_{i}(\zeta)\overline{\Psi}^{\prime}_{i}\Psi^{\prime}_{i}$ is a
huge magnitude in comparison with the vacuum energy density. It
turns out that besides the normal particle physics situations, TMT
predicts possibility of so far unknown
 states which can be realized,  for example,
in astrophysics and cosmology. Roughly speaking such exotic states
may be created if the degree of  localization of the fermion is
very small.

\subsection{The essence of the Cosmo-Particle Phenomena}

\subsubsection{Toy model I}
 To illustrate some of the properties
of such states let us start from a simplest (but idealized) model
describing the following self-consistent system: the spatially
flat FRW universe filled with the homogeneous scalar field $\phi$
and a homogeneous primordial neutrino field $N^{\prime}(t)$. The
non-canonical contribution of the primordial neutrino $N^{\prime}$
into the energy-momentum tensor reads
\begin{equation}
 T_{\mu\nu}^{(N,noncan)}=-\tilde{g}_{\mu\nu}\Lambda_{dyn}^{(N)}
 \label{Tmn-N-noncan}
\end{equation}
where
\begin{equation}
\Lambda_{dyn}^{(N)}\equiv Z_{N}(\zeta)m_{N}(\zeta)
\overline{N}^{\prime}{N}^{\prime} \label{Lambda-N}
\end{equation}
\begin{equation}
Z_{N}(\zeta)\equiv \frac{(\zeta -\zeta^{(N)}_{1})(\zeta
-\zeta^{(N)}_{2})}{2(\zeta +k)(\zeta +h_{N})}, \qquad
m_{N}(\zeta)= \frac{\mu_{N}(\zeta +h_{N})}{(\zeta +k)(\zeta
+b)^{1/2}} \label{Zeta&m-N}
\end{equation}
and $\zeta_{1,2}^{(N)}$ are defined by Eq.(\ref{zeta12}).

The neutrino has zero momenta and therefore one can rewrite
$\overline{N}^{\prime}N^{\prime}$ in the form of density
$\overline{N}^{\prime}N^{\prime}= u^{\dagger}u$ where $u$ is the
large component of the Dirac spinor $N^{\prime}$. The space
components of the 4-current
$\tilde{e}_{a}^{\mu}\overline{N}^{\prime}\gamma^{a}N^{\prime}$
equal zero. It follows from the 4-current conservation  that
$\overline{N}^{\prime}N^{\prime}= u^{\dagger}u
=\frac{const}{a^{3}}$ where $a=a(t)$ is the scale factor.
 It is convenient to rewrite the constraint
(\ref{constraint}) (adjusted to the model under consideration) in
the following form:
\begin{equation}
\frac{(b-\zeta )\left[M^{4}e^{-2\alpha\phi/M_{p}}+
V_{1}\right]-2V_{2}}{(\zeta +b)^{3/2}}=\frac{(\zeta
-\zeta^{(N)}_{1})(\zeta -\zeta^{(N)}_{2})}{(\zeta +k)^{2}}\mu_{N}
\frac{const}{a^{3}}. \label{constraint-toy}
\end{equation}

A possible solution of the constraint for the expanding universe
as the scale factor $a(t)\rightarrow\infty$ is identical to the
one studied in Sec.5 where the fermion contribution is treated as
negligible; in this case $\zeta$ is a function of
$e^{-2\alpha\phi/M_{p}}$ alone and is independent of the scale
factor.

There is however another solution where the decaying fermion
contribution $u^{\dagger}u\sim \frac{const}{a^{3}}$ to the
constraint is compensated by the appropriate behavior of the
scalar field $\zeta$. Namely if expansion of the universe is
accompanied by approaching $\zeta\rightarrow -k$ in such a way
that $(\zeta +k)^{-2} \propto a^{3}$, then the r.h.s. of
(\ref{constraint-toy}) approaches a constant. Note that the l.h.s.
of (\ref{constraint-toy}) also approaches a constant if
$\phi\rightarrow\infty$ as $a(t)\rightarrow\infty$ (recall we
assume $\alpha >0$). The described regime corresponds to a very
unexpected state of the primordial neutrino with the following
exotic features:

1. This state is different from the regular neutrino states since
$-k\neq\zeta_{1,2}$ unless a fine tuning is made.

2.  The effective mass of the neutrino in this state increases
like $(\zeta +k)^{-1}\propto a^{3/2}$ and therefore
$\rho_{(N,canon)}=T_{00}^{(N,canon)}=
m_{N}(\zeta)u^{\dagger}u\propto a^{-3/2}$.
 At the same time the dynamical neutrino $\Lambda_{dyn}^{(N)}$ term
approaches a constant: $\Lambda_{dyn}^{(N)}\propto(\zeta
+k)^{-2}u^{\dagger}u\rightarrow constant$. This means that at the
late time universe, the canonical neutrino energy density
$\rho_{(N,canon)}$
 becomes negligible in comparison with the non-canonical neutrino
 energy density $\rho_{(N,noncanon)}=T_{00}^{(N,noncanon)}=
-\Lambda_{dyn}^{(N)}$.

3. It follows immediately from Eq.(\ref{Tmn-noncan}) and item (ii)
that such cold neutrino matter possesses a pressure $p_{N}$ and
its equation of state  in the late time universe approaches the
form $p_{N}=-\rho_{N}=-\rho_{(N,canon)}$ typical for a
cosmological constant. Therefore {\it the primordial neutrino in
the described regime behaves as a sort of dark energy}.

4. In the regime $\zeta\rightarrow -k$, the scalar field $\phi$
effective potential $V_{eff}(\phi;\zeta)$, Eq.(\ref{Veff1}), and
the l.h.s. of the constraint (\ref{constraint}) have the same
order of magnitude while the r.h.s. of the constraint is
$\Lambda_{dyn}^{(N)}$. Therefore in this toy model, contributions
of the scalar field $\phi$ and the primordial neutrino into the
dark energy density are of the same order of magnitude if the
described above regime $\zeta\rightarrow -k$ is realized.

   Thus TMT
predicts a possibility of new type of states which we will refer
as {\it Cosmo-Low-Energy-Physics  (CLEP) }states.

\subsubsection{Toy Model II: Uniformly distributed non-relativistic neutrinos}

To a certain extent a more realistic model may be constructed by
studying a possible effect of uniformly distributed
non-relativistic (in the co-moving frame) neutrinos  on
cosmological scenarios in the context of TMT. In other words we
are going to study solutions motivated by what was discussed above
in the toy model I, but  now in a model where the spatially flat
FRW universe filled with the homogeneous scalar field $\phi$ and
{\it a cold gas of uniformly distributed non-relativistic
neutrinos}.

The study of the cosmology in the context of TMT becomes
complicated in the presence of the fermionic matter  because the
averaging procedure of the matter and gravity equations of motion
includes also the need to know the field $\zeta$ as a solution of
the constraint. As it was discussed in the last paragraph of
Sec.4, this generically results in the appearance of a nonlinear
$\overline{N}^{\prime}N^{\prime}$-dependence
 in all equations of motion. This makes the  procedure of
  the cosmological averaging very
unclear and perhaps practically impossible.

Significant simplifications of the described  general situation we
observed in Secs. 5 and 6 were related to the fact that in those
particular cases  the function $\zeta$ appears to be only
$\phi$-dependent or constant (or approaching constant). The
important thing we have learned in the toy model $I$ is a
possibility of the CLEP state where  $\zeta$ is close to a
constant $-k$. Recall that if the primordial neutrinos are in the
regime $\zeta\approx -k$, their masses may be very large that
justifies our choice of the gas of {\it non-relativistic}
neutrinos. On the other hand, as we have seen in the toy model,
the CLEP state can be realized if
$\overline{N}^{\prime}N^{\prime}\rightarrow 0$. A possible way to
get up such a  state might be spreading of the neutrino wave
packet during its free motion lasting a very long (of the
cosmological scale) time. However a considerable spreading of the
wave packet is again possible if neutrino is non-relativistic.

One should note however that spreading is here much more
complicated process than in the well known examples of quantum
mechanics:
 decreasing of
$\overline{N}^{\prime}N^{\prime} $ (due to the spreading) is
related to changing $\zeta$  which satisfies the nonlinear
constraint equation. So we deal in this case with nonlinear
quantum mechanics. The detailed study of  the spreading of the
wave packet in TMT is a subject of considerable interest but in
this paper we will concentrate our attention  on the consequences
of the assumption that  states with $\zeta\approx -k$  are
achievable for each of the particles of the gas of
non-relativistic primordial neutrinos. The appropriate  stage of
the universe we will call  the CLEP state because each of the
neutrinos directly participate in the cosmological expansion:
$\overline{N}^{\prime}N^{\prime}\propto a^{-3}$ and $m_{N}\propto
(\zeta +k)^{-1}\propto a^{3/2}$. We will see below that in the
context of the cosmology of the late time universe such states
provide lower energy of the universe than the states with $\zeta$
determined  in Sec.5 where there are no fermions at all. In other
words, independently of how the states with $\zeta\approx -k$ are
realized they  are energetically more preferable than states
without massive fermionic matter at all.

In the CLEP state where $\zeta(x)$ approaches a constant
($\zeta\rightarrow -k$) the procedure of the cosmological
averaging of the microscopic equations of motion becomes free of
the mentioned difficulties. For example, taking into account that
in the studied regime masses (\ref{Zeta&m-N}) of the
non-relativistic primordial neutrinos are very large, one can
ignore their kinetic energy. Then after averaging over spins of
the neutrinos, the cosmological averaging of the microscopic
non-canonical contribution to the energy-momentum tensor
$T_{\mu\nu}^{(N,can)}$ results in
\begin{equation}
<T_{\mu\nu}^{(N,can)}>_{cosm.av.}=\delta_{\mu}^{0}\delta_{\nu}^{0}
\frac{h_{N}-k}{(b-k)^{{1/2}}}\mu_{N}\frac{n}{(\zeta +k)a^{3}}+
{\cal O}\left(\frac{1}{a^{3}}\right), \label{Tmn-can-aver}
\end{equation}
where $n$ is a constant determined by the total number of the cold
neutrinos entering the CLEP regime\footnote{We  assume here that
non-relativistic primordial neutrinos in the state with
$\zeta\approx -k$ are stable and their total number is stabilized
at the late time universe.}, i.e. in the regime with $\zeta\approx
-k$. Similarly, averaging of the $\Lambda_{dyn}^{(N)}$ term gives
\begin{equation}
<\Lambda_{dyn}^{(N)}>_{cosm.av.}=(b-k)^{1/2}(h_{N}-k)\mu_{N}\frac{n}{(\zeta
+k)^{2}a^{3}}+ {\cal O}\left(\frac{1}{(\zeta +k)a^{3}}\right).
 \label{L-N-aver}
\end{equation}
Hence the appropriate averaged expression of the microscopic
non-canonical contribution to the energy-momentum tensor
$T_{\mu\nu}^{(N,noncan)}$, Eq.(\ref{Tmn-N-noncan}), is then
\begin{equation}
<T_{\mu\nu}^{(N,noncan)}>_{cosm.av.}=
-\tilde{g}_{\mu\nu}<\Lambda_{dyn}^{(N)}>_{cosm.av.}
\label{Tmn-noncan-aver}
\end{equation}

Taking into account the homogeneity of the scalar field $\phi$ and
Eq.(\ref{L-N-aver}), the result of the cosmological averaging of
the constraint (\ref{constraint-toy}) can be represented in the
form
\begin{eqnarray}
\frac{(b+k)\left(M^{4}e^{-2\alpha\phi/M_{p}}+V_{1}\right)
-2V_{2}}{(b-k)^{2}} \nonumber
\\
=(b-k)^{1/2}(h_{N}-k)&&\mu_{N}\frac{n}{(\zeta +k)^{2}a^{3}}+{\cal
O}\left(\zeta +k\right)
 \label{constr-k-av}
\end{eqnarray}

Similar to the toy model of Sec.7.1.1, the constraint
(\ref{constr-k-av}) allows a "CLEP solution" where the decay of
the neutrino density with the expansion of the universe is
accompanied by approaching $\zeta\rightarrow -k$ in such a way
that $(\zeta +k)^{-2} \propto a^{3}$, and then the r.h.s. of
(\ref{constr-k-av}) as well as $<\Lambda_{dyn}^{(N)}>_{cosm.av.}$
in  Eq.(\ref{Tmn-noncan-aver}) approach  constants. At the same
time, $<T_{\mu\nu}^{(N,can)}>_{cosm.av.}$,
Eq.(\ref{Tmn-can-aver}), approaches zero. Therefore in the CLEP
regime, the neutrino energy-momentum tensor
$<T_{\mu\nu}^{(N)}>_{cosm.av.}$ is reduced to the approaching
constant (as $a(t)\rightarrow\infty$) non-canonical part of the
neutrino energy-momentum tensor
\begin{equation}
<T_{\mu\nu}^{(N)}>_{cosm.av.}= -\tilde{g}_{\mu\nu}
(b-k)^{1/2}(h_{N}-k)\mu_{N}\frac{n}{(\zeta +k)^{2}a^{3}}+{\cal
O}\left(\zeta +k\right) \label{Tmn-noncan-aver-asympt}
\end{equation}

Now we want to obtain the homogeneous scalar field $\phi(t)$
contribution $T_{\mu\nu}^{(\phi)}$ into the energy-momentum tensor
in the CLEP regime of our model, i.e. when due to the balance
dictated by the constraint, uniformly distributed non-relativistic
neutrinos demand that the averaged value of $\zeta$ approaches
$-k$. Comparing the latter with the expression for $\zeta
=\zeta_{0}$ in the case of the absence of fermion matter,
Eq.(\ref{zeta-without-ferm}), we conclude that the presence of the
cold neutrino gas in the CLEP regime essentially changes
$V_{eff}(\phi;\zeta)$, Eq.(\ref{Veff1}). In fact, instead of
$V_{eff}^{(0)}(\phi)$ as $\zeta =\zeta_{0}$, Eq.(\ref{Veffvac}),
we obtain in the case $\zeta\rightarrow -k$
\begin{equation}
V_{eff}^{(CLEP)}(\phi)\equiv
V_{eff}(\phi;\zeta)|_{\zeta\rightarrow -k}=
\frac{b\left(M^{4}e^{-2\alpha\phi/M_{p}}+V_{1}\right)
-V_{2}}{(b-k)^{2}}+{\cal O}\left(\zeta +k\right). \label{Veff3}
\end{equation}

 We would like to stress that  this reorganization of the
scalar field $\phi$ dynamics in the presence of the cold neutrino
gas in the CLEP regime is the direct result of a general feature
of TMT following from the constraint (\ref{constraint}): in the
space-time region occupied by a fermion, the scalar field $\zeta$
provides a certain balance between the energy densities of the
scalar field $\phi$ and fermion.

It is very interesting  that in the CLEP regime the role of the
constraint becomes still more important. So far we have discussed
separately the neutrino contribution into the total
energy-momentum tensor, Eq.(\ref{Tmn-noncan-aver-asympt}), and the
homogeneous scalar field $\phi$ contribution which can be written
 in a concise  form
\begin{equation}
T_{\mu\nu}^{(\phi)}=\phi_{,\mu}\phi_{,\nu}-\frac{1}{2}
\tilde{g}_{\mu\nu}\tilde{g}^{\alpha\beta}\phi_{,\alpha}\phi_{,\beta}
+\tilde{g}_{\mu\nu}V_{eff}^{(CLEP)}(\phi). \label{Tmn-phi}
\end{equation}
However due to the constraint (\ref{constr-k-av}), the separation
 of the neutrino and scalar field $\phi$ contributions
into the total energy-momentum tensor written in the form
\begin{equation}
<T_{\mu\nu}^{(tot)}>_{cosm.av.}=<T_{\mu\nu}^{(N)}>_{cosm.av.}
+T_{\mu\nu}^{(\phi)} \label{Tmn-total-aver-asympt}
\end{equation}
loses clarity in the CLEP regime\footnote{Recall that the
constraint
  is just the consistency
condition of the equations of motion.}. In fact, {\it using the
constraint} (\ref{constr-k-av}) {\it one can represent "the
neutrino contribution"} $<T_{\mu\nu}^{(N)}>_{cosm.av.}$ {\it into}
$<T_{\mu\nu}^{(tot)}>_{cosm.av.}$ {\it in terms of the scalar
field} $\phi$ {\it alone} and thus the total energy and pressure
in the CLEP state can be written in an equivalent form where they
are only $\phi$-dependent:
\begin{equation}
\rho_{tot}\equiv <T_{00}^{(tot)}>_{cosm.av.}=
\frac{1}{2}\dot{\phi}^{2}+U_{eff}^{(tot)}(\phi) \label{rho-tot}
\end{equation}
\begin{equation}
p_{tot}\equiv <T_{ii}^{(tot)}>_{cosm.av.}=
\frac{1}{2}\dot{\phi}^{2}-U_{eff}^{(tot)}(\phi). \label{p-tot}
\end{equation}
Here the effective potential $U_{eff}^{(tot)}(\phi)$ is the sum
\begin{equation}
U_{eff}^{(tot)}(\phi)\equiv \Lambda +V_{q-like}(\phi),
\label{Ueff-phi}
\end{equation}
of the effective cosmological constant
\begin{equation}
\Lambda = \frac{V_{2}-kV_{1}}{(b-k)^{2}} \label{Lambda-nu}
\end{equation}
and the potential
\begin{equation}
V_{q-like}(\phi)=
-\frac{k}{(b-k)^{2}}M^{4}e^{-2\alpha\phi/M_{p}}+{\cal
O}\left(\zeta +k\right). \label{pot-nu}
\end{equation}
$\Lambda$ and $V_{q-like}(\phi)$ are positive if
\begin{equation}
V_{2}-kV_{1}>0 \quad and \quad k<0 \label{param-CLEP}
\end{equation}
that we will assume in what follows. Recall also the conditions
(\ref{param-vac}) and (\ref{bv1>2v2}).

Thus, similar to the absence of fermions case discussed in Sec.5,
the evolution of the late time universe filled with the
homogeneous scalar field $\phi$ {\it and the cold gas of uniformly
distributed non-relativistic neutrinos in the state with}
$\zeta\approx -k$ {proceeds  as it would be in the standard field
theory model} (non-TMT) including  both the cosmological constant
 and the scalar field $\phi$ with the exponential
potential {\it but without fermions}. In other words the role of
the gas of uniformly distributed non-relativistic neutrinos in the
CLEP state consists only in the change of the dark energy in
comparison with the absence of fermions case of Sec.5. Namely, the
dark energy in the CLEP regime is also associated only with  the
homogeneous scalar field $\phi$ (as if there were no neutrinos)
but now with the effective potential $U_{eff}^{(tot)}(\phi)$
instead of $V_{eff}^{(0)}(\phi)$, Eq.(\ref{Veffvac}), obtained for
the universe filled only with the homogeneous scalar field $\phi$.

 The {\it remarkable result} consists
in the fact that
\begin{equation}
V_{eff}^{(0)}(\phi)-U_{eff}^{(tot)}(\phi)\equiv
\frac{\left[(b+k)\left(V_{1}+M^{4}e^{-2\alpha\phi /M_{p}}\right)
-2V_{2}\right]^{2}}
{4(b-k)^{2}\left[b\left(V_{1}+M^{4}e^{-2\alpha\phi /M_{p}}\right)
-V_{2}\right]}>0. \label{L-L0}
\end{equation}
where the conditions (\ref{param-vac}) and (\ref{bv1>2v2}) have
been used. This means that {\it the universe with the gas of
uniformly distributed non-relativistic neutrinos in the CLEP state
is energetically more preferable than the one in the absence of
fermions case}.

\subsection{Cosmological Solutions for the Late Time Universe
 in the CLEP Regime}

  Cosmological equations for a spatially
 flat FRW universe filled with the homogeneous scalar field $\phi$ and
    the cold gas
of uniformly distributed non-relativistic neutrinos with wave
function $N^{\prime}(t)$ in the regime with $\zeta\approx -k$ are
the following
\begin{equation}
\left(\frac{\dot{a}}{a}\right)^{2}=\frac{1}{3M_{p}^{2}}\rho_{tot}
 \label{FRW-eq1}
\end{equation}
\begin{equation}
\ddot{\phi}+3\frac{\dot{a}}{a}\dot{\phi} +\frac{2\alpha
k}{(b-k)^{2}M_{p}}M^{4}e^{-2\alpha\phi/M_{p}} +{\cal
O}\left((\zeta +k)e^{-2\alpha\phi/M_{p}}\right)=0,
\label{phi-eq-cosm}
\end{equation}
\begin{equation}
i\gamma^{0}\left(\partial_t
+\frac{3}{2}\frac{\dot{a}}{a}\right)N^{\prime}(t)-m(\zeta)N^{\prime}(t)=0
\label{neutrino-eq}
\end{equation}
Here $\rho_{tot}$ is given by Eq.(\ref{rho-tot}). The simplest way
to obtain Eq.(\ref{phi-eq-cosm}) is to use the scalar field $\phi$
equation in the form given by Eq.(\ref{phi-constr}). In
Eq.(\ref{neutrino-eq}) for the neutrino in the spatially
 flat FRW universe we have taken into account that
 the fermionic contribution to the spin-connection (\ref{A7})
 turns into zero in the rest frame of the fermion.

The solution of the neutrino equation is straightforward
\begin{equation}
N^{\prime}(t) =\frac{C}{a^{3/2}}
\begin{pmatrix} 1 \\
 0\\
 0\\
 0
 \end{pmatrix}\cdot e^{-i\int
m\left(\zeta(t)\right)dt}
\end{equation}
where $m\left(\zeta(t)\right)$ is defined by the second equation
in (\ref{Zeta&m-N}) and $C$ is a constant.

In the CLEP regime $\zeta +k\propto a^{-3/2}$ and therefore the
correction term ${\cal O}\left((\zeta
+k)e^{-2\alpha\phi/M_{p}}\right)$ in (\ref{phi-eq-cosm}) becomes
negligible at the late time. The effective potential
$U_{eff}^{(tot)}(\phi)$ monotonically decays to the effective
cosmological constant $\Lambda$, Eq.(\ref{Lambda-nu}), as
$\phi\rightarrow\infty$. Hence for the class of solutions where
$\phi\rightarrow\infty$ and $\dot{\phi}\rightarrow 0$ as
$a(t)\rightarrow\infty$, the asymptotic behavior of the universe
is governed by the effective cosmological constant $\Lambda$.

One should note that the needed smallness of $\Lambda$ cannot be
here explained by means of a see-saw mechanism in spite of the
fact that $\Lambda$ is bounded from above(\ref{L-L0}) by
$\Lambda^{(0)}$ for which a see-saw mechanism was possible (see
the end of Sec.5, item (i)). In fact, if we want to realize a
similar situation in the case $V_2<0$ and $-kV_1<|V_2|$ then we
obtain $\Lambda <0$ which is of course unsatisfactory result.

However if $V_2>0$ or alternatively $V_2<0$ and $-kV_1>|V_2|$,
 the needed smallness of $\Lambda$ is achieved again through a choice
of huge value(s) of the dimensionless parameter(s) $b$ and/or $k$,
similar to what was discussed in item (ii) at the end of Sec.5.

One more possibility is to get $\Lambda$ very small by tuning the
parameters such that $V_{2}-kV_{1}$ is very small. If
$V_{2}-kV_{1}=0$ then $\phi$ becomes the regular quintessence
field with an exponential potential\footnote{For some details of
the model with $V_{1}=V_{2}=0$ see Appendix.}.

For illustration of what kind of solutions one can expect, let us
take the {\em particular value} for the parameter $\alpha$, namely
$\alpha =\sqrt{3/8}$. Then for the late time universe in the CLEP
regime, when $\zeta$ is close enough to $-k$,
 the cosmological equations allow the following analytic solution:
 \begin{equation}
\phi(t)=\frac{M_{p}}{2\alpha}\varphi_{0}+
\frac{M_{p}}{2\alpha}\ln(M_{p}t), \qquad a(t)\propto
t^{1/3}e^{\lambda t}, \label{a-sol-nu}
\end{equation}
where
\begin{equation}
\lambda =\frac{1}{M_{p}}\sqrt{\frac{\Lambda}{3}}, \qquad
e^{-\varphi_{0}}=
\frac{2(b-k)^{2}M_{p}^{2}}{\sqrt{3}|k|M^{4}}\sqrt{\Lambda}.
\label{phi-0}
\end{equation}
and $\Lambda$ is determined by Eq.(\ref{Lambda-nu}). The mass of
the neutrino in such CLEP state increases exponentially in time
and its $\phi$ dependence is double-exponential:
\begin{equation}
m_{N}|_{clep}\sim (\zeta +k)^{-1}\sim a^{3/2}(t)\sim
t^{1/2}e^{\frac{3}{2}\lambda t}\sim \exp\left[\frac{3\lambda
e^{-\varphi_{0}}}{2M_{p}} \exp\left(\frac{2\alpha}{M_{p}}\phi
\right)\right]. \label{m-t-phi}
\end{equation}

\section{Conclusions and Discussion}
\subsection{Main Results}
In this paper we have shown that the features of a fermion in TMT
depend on its density and therefore states of the fermion (in this
generalized sense)  may be very much different from what is known
about regular fermions. This is the reason why generically we
should start in TMT from a so-called primordial fermion. Only in
the normal particle physics conditions, i.e. when the fermion
energy density has huge values in comparison with the vacuum
energy density, the primordial fermion behaves as the regular
fermion. Both the decoupling of fermion from the quintessence-like
scalar field and restoration of Einstein equations occur also in
this regime.

It turns out that besides states of the regular fermions, the
primordial fermion can be in exotic states which might be of
interest in cosmology and astrophysics\footnote{Exotic fermion
fields of different nature has been studied in Ref.\cite{ahlu}.}.
In particular, for states of the primordial fermion with energy
density close to the dark energy density, TMT predicts very
interesting properties, like for example: the non-canonical
(proportional to the metric tensor) term of the energy-momentum
tensor becomes important; the mass of the fermion depends on its
density, etc.. We call such states Cosmo-Low-Energy-Physics (CLEP)
states.

At this stage of research we have found out  the solution in the
CLEP regime for the model of the spatially flat FRW universe
filled with the homogeneous scalar field $\phi$ and the cold gas
of uniformly distributed non-relativistic neutrinos. The main
features of the CLEP regime are the following:
\begin{itemize}

\item  In the CLEP state, i.e. as the neutrino density approaches zero and
$\zeta\rightarrow -k$, the canonical contribution of the uniformly
distributed non-relativistic neutrinos to the energy-momentum
tensor becomes negligible in comparison with their non-canonical
(proportional to the metric tensor) contribution.

\item Due to the constraint (\ref{constraint}), the above neutrino
contribution to the energy-momentum tensor can be expressed in
term of the scalar field $\phi$ alone. Hence the gas of the
uniformly distributed non-relativistic neutrinos in the CLEP state
behaves as a genuine dark energy.

\item The original scalar field $\phi$ effective potential is also
 $\zeta$ dependent and thus the CLEP
state of the uniformly distributed non-relativistic neutrinos
influences in a fundamental way the scalar field $\phi$
contribution to the dark energy.

\item The late time universe in the CLEP regime is governed by
both a cosmological constant and a decaying exponential potential.

\item The mass of
the neutrino in the CLEP state increases according to
\begin{equation}
m_{N}|_{clep}\propto
\left(\overline{N}^{\prime}N^{\prime}\right)^{-1/2}\propto
a^{3/2}(t)\label{m-dens}
\end{equation}

\end{itemize}

The latter feature allows to make a crude estimation for the
present day mass of the CLEP neutrinos. It is clear that
transition of cold neutrinos to the CLEP regime can occur after
their decoupling, i.e. at the temperature $\sim 0.1 MeV$. Hence
the ratio of the present scale factor $a_{0}$ to the scale factor
at the neutrino decoupling epoch $a_{dec}$ is $a_{0}/a_{dec}\sim
10^{9}$. Assume that the neutrino mass right before entering the
CLEP regime is not less than the mass of the regular neutrino.
Then using Eq.(\ref{m-dens}) one can estimate
$m_{N}|_{clep,present}\sim 10^{14}m_{\nu}$ where $m_{\nu}$ is of
the typical order of a mass of a regular neutrino. For
$m_{\nu}\sim (10^{-2}-1)eV$ it gives the following estimation for
masses of the CLEP state neutrinos at the present day universe as
a result of the cosmological expansion:
\begin{equation}
m_{N}|_{clep}\sim (1-10^{2})TeV. \label{m-int}
\end{equation}

One can obtain also a crude estimation of a typical size of the
space region where the wave function of the present CLEP neutrino
state is non-zero. Taking the mass parameter $m_{\nu}\sim
10^{-2}eV$ right before entering the CLEP regime, we estimate that
the linear size of the neutrino at this stage is larger than its
Compton wave length $\sim 10^{-3}cm$. Then the cosmological
expansion results in the linear size of the support of the wave
function of the present CLEP neutrino state to be larger than
$10^{6}cm$.

 We did not study yet the transition of the regular
neutrinos (which constitute at least a fraction of the dark
matter) into the CLEP state (where neutrinos behave as a dark
energy). We expect that a possible mechanism responsible for
initiating this transition is spreading of the regular neutrino
wave packet that has an effect the decreasing of the neutrino
density; the latter is a necessary element of the CLEP regime. Due
to the constraint (\ref{constraint}), the description of this
spreading requires the consideration of a non linear quantum
mechanics which makes the complete solution of this problem
complicated enough.

Nevertheless our statement is that the described CLEP regime is
favorable from the energetic point of view. In fact, from the two
possibilities existing as
$\overline{N}^{\prime}N^{\prime}\rightarrow 0$ (one is the normal
one with the neutrino energy density decreasing to zero; another -
the CLEP regime), the CLEP state provides the lowest {\it total}
energy density.

The possibility of the CLEP state does not require tuning of the
dimensionless parameters $b$, $k$, $h_{N}$ and values of $V_{1}$
and $V_{2}$ in the action (\ref{totaction}) as well as of the
value of the (positive) integration constant $M^{4}$ in
Eq.(\ref{varphi}). The effect is also not sensitive to the value
of $n$ determined by the total number of cold neutrinos entering
the CLEP regime. However the scale symmetry (\ref{stferm})
(reduced in the Einstein frame to the shift symmetry
(\ref{phiconst})) and its spontaneous breakdown by means of
Eq.(\ref{varphi}) have a decisive role in the structure of the
{\it dynamically generated effective potential}.

It is also very important that the gauge couplings of the neutrino
that may appear in more realistic models, does not affect the
possibility of the CLEP state , i.e. we are at this point  free to
use either the active neutrino with non-abelian gauge interactions
or a sterile (non standard model) neutrino.

It is interesting to compare the CLEP regime solution of Sec.7 (in
the context of TMT) with the results of  Sec.2 where possibilities
of fermion contributions to the dark energy have been demonstrated
in the framework of regular (non TMT) toy models. In these non-TMT
models, contributions to the energy-momentum tensor proportional
to the metric tensor are really generated but there are no reasons
for the proportionality factors to approach non zero constant
values while this occurs in the CLEP regime.

\pagebreak

\subsection{Outlook}
Finally we would like to outline few possible further developments
of the CLEP regime solutions.

\begin{itemize}

\item {\it Interactions of the CLEP State Neutrinos and Cosmic
Rays.}

$\overline{N}^{\prime}N^{\prime}$ which
  is
 a probability density for
 non-relativistic neutrino in the co-moving frame, is anomalously
 small in the CLEP state:
 $(\overline{N}^{\prime}N^{\prime})|_{clep}\propto a^{-3}(t)$.
 Therefore it is very hard to observe the primordial
neutrinos in the CLEP states. Furthermore, an attempt to detect
the fermion in the CLEP state should be accompanied by its
localization, i.e developing a large
$\overline{N}^{\prime}N^{\prime}$ that destroys the condition for
the existence of the CLEP state.

Using our estimation for masses of the CLEP state neutrinos at the
present day universe (\ref{m-int}) and taking into account their
{\it standard} weak interaction, one can speculate about
possibility of {\it decays of the CLEP state neutrinos}. Smallness
of the CLEP neutrino wave function leads to a very strong damping
of the amplitude of the decay. However, propagator of the gauge
boson exhibits a {\it resonance enhancement} of the amplitude of
the decay of the the neutrino in the CLEP state $N_{clep}$ when
the energy $E_{1}$ of the charged lepton $l^{-}$ produced in the
vertex $N_{clep}l^{-}W^{+}$ (or $N_{clep}\nu Z$) satisfies
$E_{1}\approx (m^{2}_{N}|_{clep}-M_{W}^{2})/(2m_{N}|_{clep})$.
This becomes possible when the mass of the CLEP state neutrinos
$m_{N}|_{clep}$ is larger than masses of the gauge bosons
$W^{\pm}$ and $Z$. In addition, the width of such weak decays of
$N_{clep}$ into three particles is enhanced because the phase
space is proportional to $\left(m_{N}|_{clep}\right)^{5}$.
Estimations (\ref{m-int}) for masses of the CLEP state neutrinos
at the present day universe allow to expect that decays of the
CLEP state neutrinos might be the origin of TeV cosmic rays. These
ideas and preliminary estimations show that  decays of the CLEP
neutrino (both with masses below $M_{W},M_{Z}$, producing low
energy cosmic rays, and with masses above $M_{W},M_{Z}$, producing
TeV and even higher energy cosmic rays) deserve detailed study. In
this sense the CLEP states exhibit  simultaneously  new physics at
very low densities and for very high particle masses.

\item {\it Domain Structure of the Universe.}
 Existence of the CLEP states suggests the possibility of a domain
structure of the dark energy. In fact, if for some reasons there
are space-time regions empty of fermionic matter then the dark
energy in such regions is governed by the scalar field with the
potential (\ref{Veffvac}). Let us call such regions as type $I$
regions. As we have seen, the energy density of the type $I$
regions is bigger than the energy density in the regions with CLEP
state neutrinos (see Eq.(\ref{L-L0})). Let us call the latter as
type $II$ regions. One can think of a typical situation when the
bubble of the type $II$ region is situated in the midst of the
type $I$ region.  Then
 the difference of the pressures between the regions {\it I} and {\it II}
 causes the CLEP bubble to expand replacing the
 vacuum of the region {\it I} by the CLEP state. Due to this {\it mechanical} effect
 the size of the CLEP bubble should expand faster than the
 universe itself. Therefore the above estimations for the mass,
 Eq. (\ref{m-int}), and size (see paragraph after Eq. (\ref{m-int}))
 of the present CLEP neutrino resulting from the cosmological
 expansion
 could be enhanced due to this  mechanical effect.

 \item The difference between the dark energy states in regions $I$
and $II$ is related to the different values of the scalar $\zeta$
which is determined by the constraint(\ref{constraint}). It is
interesting that the electromagnetic field does not enter the
constraint. Therefore, the dark energy bubbles including walls
between them are transparent for radiation.

 \item {\it Super-Acceleration as a Possible Consequence of
  Decays of the CLEP State Neutrinos.}
  Let us consider a model of the present day universe containing both
  the regions {\it I} and the regions {\it II} and assume that
  the cosmologically averaged dark energy density $\rho_{d.e.}$
  is a dominant fraction. The magnitude of $\rho_{d.e.}$
  depends on the number of the regions {\it II}, i.e. on the number
  of the CLEP state neutrinos. Therefore the possibility of the
   discussed above  decays  of  the CLEP state neutrinos
  may result in a growth of  $\rho_{d.e.}$ that exhibits itself
  via super-acceleration of the universe. It is interesting that
these decays may become non negligible due to the mentioned above
resonance enhancement of the amplitude {\it only  at a late enough
epoch} since the mass of the CLEP state neutrino should grow above
masses of the gauge bosons. Hence one can expect a certain
correlation between the cosmic rays physics on the one hand and
the dark energy physics on the other hand.

\item {\it Inhomogeneous CLEP States and Dark Matter.}
The research presented in this paper is {\it the first step in
studying the CLEP states}: here we have restricted ourselves in
effects of the CLEP state in homogeneous cosmology. Possible
inhomogeneous (local) effects of the CLEP states and their
connection with dark matter will be a subject of future research.

\end{itemize}

\section*{Acknowledgments}

We thank S.P. de Alwis, L. Amendola, S. Ansoldi, J. Bekenstein, S.
del Campo, A. Dolgov, V. Elias, P. Frampton, K. Freese, S.B.
Giddings, P. Gondolo, J.B. Hartle, B-L. Hu, G. Huey, P.Q. Hung, D.
Kazanas, A. Kheyfets, D.G. McKeon,   D. Marolf, J. Morris, M.
Milgrom, V. Miransky, A. Nelson, H. Nielsen, Y.Jack Ng, E.
Nissimov, S. Nussinov, S. Pacheva, L. Parker, R. Peccei, M.
Pietroni, T. Piran, R. Rosenfeld, A. Roura, V. Rubakov, E.
Spallucci, A. Vilenkin, I. Waga, P. Wesson  and A. Zee for helpful
conversations on different stages of this research. One of us
(E.G.) wants to thank the universities of Trieste and Udine, the
Michigan Center for Theoretical Physics and ICRA (Pescara and
Rome) for hospitality.
\appendix

\section{Connection in the original and Einstein frames}

We present here what is the dependence of the spin connection
$\omega_{\mu}^{ab}$ on $e^{a}_{\mu}$, $\zeta$, $\Psi$ and
$\overline{\Psi}$. Varying the action (\ref{totaction}) with
respect to $\omega_{\mu}^{ab}$ and making use that
\begin{equation}
R(V,\omega)\equiv
-\frac{1}{4\sqrt{-g}}\varepsilon^{\mu\nu\alpha\beta}\varepsilon_{abcd}
e^{c}_{\alpha}e^{d}_{\beta}R_{\mu\nu}^{ab}(\omega) \label{A1}
 \end{equation}
we obtain
\begin{eqnarray}
&&\varepsilon^{\mu\nu\alpha\beta}\varepsilon_{abcd}\left[(\zeta
+b_{g}) e^{c}_{\alpha}D_{\nu}e^{d}_{\beta}
+\frac{1}{2}e^{c}_{\alpha}e^{d}_{\beta}\left(\zeta_{,\nu}+\frac{\alpha}{M_{p}}\phi_{,\nu}\right)\right]
\nonumber\\
&+& \frac{\kappa}{4}\sqrt{-g}(\zeta
+k)e^{c\mu}\varepsilon_{abcd}\overline{\Psi}
\gamma^{5}\gamma^{d}\Psi=0, \label{A2}
 \end{eqnarray}
where
\begin{equation}
D_{\nu}e_{a\beta}\equiv\partial_{\nu}e_{a\beta} +\omega_{\nu
a}^{d}e_{d\beta} \label{A3}
 \end{equation}
The solution of Eq. (\ref{A2}) is represented in the form
\begin{equation}
\omega_{\mu}^{ab}=\omega_{\mu}^{ab}(e)   +
K_{\mu}^{ab}(e,\overline{\Psi},\Psi) + K_{\mu}^{ab}(\zeta, \phi)
\label{A4}
 \end{equation}
where
\begin{equation}
\omega_{\mu}^{ab}(e)=e_{\alpha}^{a}e^{b\nu}\{
^{\alpha}_{\mu\nu}\}- e^{b\nu}\partial_{\mu}e_{\nu}^{a} \label{A5}
 \end{equation}
is the Riemannian part of the spin-connection,
\begin{equation}
K_{\mu}^{ab}(e,\overline{\Psi},\Psi)= \frac{\kappa}{8}\frac{\zeta
+k}{\zeta +b_{g}}
\eta_{cn}e_{d\mu}\varepsilon^{abcd}\overline{\Psi}
\gamma^{5}\gamma^{n}\Psi \label{A7}
 \end{equation}
is the fermionic contribution that differs from the familiar
one\cite{Gasperini} by the factor $\frac{\zeta +k}{\zeta +b_{g}}$
and
\begin{equation}
K_{\mu}^{ab}(\zeta, \phi)=\frac{1}{2(\zeta
+b_{g})}\left(\zeta_{,\alpha}+\frac{\alpha}{M_{p}}\phi_{,\alpha}\right)(e_{\mu}^{a}e^{b\alpha}-
e_{\mu}^{b}e^{a\alpha}) \label{A6}
 \end{equation}
is  the non-Riemannian part of the spin-connection originated by
specific features of TMT.

In the Einstein frame, i.e. in terms of variables defined by
Eq.(\ref{ctferm}), the spin-connection read
\begin{equation}
\omega^{\prime ab}_{\mu}=\omega^{ab}_{\mu}(\tilde{e}) +
\frac{\kappa}{8}
\eta_{cn}\tilde{e}_{d\mu}\varepsilon^{abcd}\overline{\Psi}^{\prime}
\gamma^{5}\gamma^{n}\Psi^{\prime} \label{A7}
 \end{equation}
 which is exactly the spin-connection of the Einstein-Cartan
 space-time\cite{Gasperini} with the vierbein $\tilde{e}^{a}_{\mu}$.

\section{The CLEP Regime Cosmological Solutions in a Model
 Without Explicit Potentials}

In Ref. \cite{GK4} we have studied scalar field $\phi$ cosmology
(without fermions) in a model without potentials in the action,
i.e. $V_{1}=V_{2}=0$ in (\ref{totaction}). In the framework of the
models with $V_{1}=V_{2}=0$, in this Appendix we are going  to
discuss cosmological solutions in the CLEP regime.

 In the model with $V_{1}=V_{2}=0$, the constraint (\ref{constraint}) is
reduced to the following:
\begin{equation}
\frac{(b-\zeta)}{(\zeta +b)^{2}}M^{4}e^{-2\alpha\phi/M_{p}}=
\Lambda_{dyn}^{(ferm)} \label{app1}
\end{equation}

{\bf A.} {\it The universe filled with the homogeneous scalar
field $\phi$ alone.}

Instead of the results of Sec.5 we obtain:
\begin{itemize}

\item
the constraint (\ref{app1}) yields $\zeta =b$;

\item
the cosmological constant $\Lambda^{(0)}=0$

\item
 the scalar field effective potential
\begin{equation}
V_{eff}^{(0)}(\phi)|_{V_{1}=V_{2}=0}=
\frac{M^{4}}{4b}e^{-2\alpha\phi/M_{p}} \label{app2}
\end{equation}
is generated {\it only} through  the spontaneous breakdown of the
shift symmetry (\ref{phiconst}) by means of Eq.(\ref{varphi}).

\end{itemize}

 {\bf B.} {\it The universe filled with the homogeneous scalar field $\phi$
and a cold gas of uniformly distributed non-relativistic
neutrinos.}

In the CLEP regime, i.e. as $\zeta\rightarrow -k$,
Eqs.(\ref{Tmn-can-aver}), (\ref{L-N-aver}) and
(\ref{Tmn-noncan-aver})  hold as well. However, instead of
Eq.(\ref{constr-k-av}), the averaged constraint reads now
\begin{equation}
\frac{b+k}{(b-k)^{2}}M^{4}e^{-2\alpha\phi/M_{p}} \nonumber\\
=(b-k)^{1/2}(h_{N}-k)\mu_{N}\frac{n}{(\zeta +k)^{2}a^{3}}+{\cal
O}\left((\zeta +k)e^{-2\alpha\phi/M_{p}}\right).
 \label{app3}
\end{equation}

Due to the constraint (\ref{app3}) the correction to the main term
of $<T_{\mu\nu}^{(N)}>_{cosm.av.}$,
Eq.(\ref{Tmn-noncan-aver-asympt}), is now different:
\begin{equation}
<T_{\mu\nu}^{(N)}>_{cosm.av.}= -\tilde{g}_{\mu\nu}
(b-k)^{1/2}(h_{N}-k)\mu_{N}\frac{n}{(\zeta +k)^{2}a^{3}}+{\cal
O}\left((\zeta +k)e^{-2\alpha\phi/M_{p}}\right). \label{app4}
\end{equation}
Similar change of the correction term to $V_{eff}^{(CLEP)}(\phi)$,
Eq.(\ref{Veff3}), takes place now:
\begin{equation}
V_{eff}^{(CLEP)}(\phi)=
\frac{b\left(M^{4}e^{-2\alpha\phi/M_{p}}+V_{1}\right)
-V_{2}}{(b-k)^{2}}+{\cal O}\left((\zeta
+k)e^{-2\alpha\phi/M_{p}}\right). \label{app5}
\end{equation}

Keeping the definitions (\ref{Tmn-phi}) and
(\ref{Tmn-total-aver-asympt}) and using the constraint
(\ref{app3}) one can again represent the neutrino contribution
into $<T_{\mu\nu}^{(tot)}>_{cosm.av.}$  in terms of the scalar
field $\phi$  alone and thus the total energy and pressure in the
CLEP regime can be written in an equivalent form given by
Eqs.(\ref{rho-tot}) and (\ref{p-tot}) where now the cosmological
constant equals zero and $U_{eff}^{(tot)}(\phi)$ reads
\begin{equation}
U_{eff}^{(tot)}(\phi)\equiv
-\frac{k}{(b-k)^{2}}M^{4}e^{-2\alpha\phi/M_{p}}+{\cal
O}\left((\zeta +k)e^{-2\alpha\phi/M_{p}}\right). \label{app6}
\end{equation}

Using Eqs.(\ref{app2}) and (\ref{app6}) we obtain the result
similar to Eq.(\ref{L-L0}):
\begin{equation}
V_{eff}^{(0)}(\phi)-U_{eff}^{(tot)}(\phi)\equiv \frac{(b+k)^{2}}
{4b(b-k)^{2}}M^{4}e^{-2\alpha\phi /M_{p}}>0. \label{app7}
\end{equation}

 Cosmological equations for a spatially
 flat FRW universe filled with the homogeneous scalar field $\phi$ and
    the cold gas
of uniformly distributed non-relativistic neutrinos in the state
with $\zeta\rightarrow -k$ coincide in the form with
Eqs.(\ref{FRW-eq1}) and (\ref{phi-eq-cosm}) where $\rho_{tot}$ is
given by Eq.(\ref{rho-tot}) with $U_{eff}^{(tot)}(\phi)$ given by
Eq.(\ref{app6}). Neglecting the corrections $\propto(\zeta
+k)e^{-2\alpha\phi/M_{p}}$ we deal with the quintessence field
with an exponential potential. The appropriate  solution is:
\begin{equation}
\phi(t)=\frac{M_{p}}{2\alpha}\varphi_{0}+
\frac{M_{p}}{\alpha}\ln(M_{p}t);  \qquad a\propto
t^{1/2\alpha^{2}} \label{app8},
\end{equation}
where
\begin{equation}
e^{-\varphi_{0}}=\frac{(b-k)^{2}}{4|k|\alpha^{4}}(3-2\alpha^{2})
\left(\frac{M_{p}}{M}\right)^{4}. \label{app9}
\end{equation}
This solution describes an accelerating expansion if
$\alpha^{2}<\frac{1}{2}$. It is easy to check that in this case
the corrections $\propto(\zeta +k)e^{-2\alpha\phi/M_{p}}$ we have
neglected, approach zero faster than the main terms of the
cosmological equations and the mass of the neutrino in such CLEP
regime increases  in time as
\begin{equation}
m_{N}|_{CLEP}\sim (\zeta +k)^{-1}\sim t^{\frac{3}{4\alpha^{2}}-1}
>t^{1/2}.\label{app10}
\end{equation}

\end{document}